\documentclass{aa}

\pdfoutput=1
\usepackage{txfonts}
\usepackage{graphicx}
\usepackage{natbib}
\bibpunct{(}{)}{;}{a}{}{,}

\begin{document}

\title{The ALHAMBRA survey: Accurate merger fractions by PDF analysis of photometric close pairs\thanks{Based on observations collected at the German-Spanish Astronomical Center, Calar Alto, jointly operated by the Max-Planck-Institut f\"ur Astronomie (MPIA) at Heidelberg and the Instituto de Astrof\'{\i}sica de Andaluc\'{\i}a (CSIC)}}

\author{C.~L\'opez-Sanjuan\inst{1}
\and A.~J.~Cenarro              \inst{1}    
\and J.~Varela                  \inst{1}    
\and K.~Viironen		 \inst{1}   
\and A.~Molino			 \inst{2}    
\and N.~Ben\'{\i}tez		 \inst{2}    
\and P.~Arnalte-Mur		 \inst{3}    
\and B.~Ascaso			\inst{4,2}  
\and L.~A.~D\'{\i}az-Garc\'{\i}a	\inst{1}
\and A.~Fern\'andez-Soto	\inst{5,6}
\and Y.~Jim\'enez-Teja		\inst{7}
\and I.~M\'arquez		\inst{2}
\and J.~Masegosa		\inst{2}
\and M.~Moles			\inst{1,2}
\and M.~Povi\'c			\inst{2}  
\and J.~A.~L.~Aguerri		\inst{8,9}   
\and E.~Alfaro			\inst{2}
\and T.~Aparicio-Villegas	\inst{7,2}     
\and T.~Broadhurst		\inst{10}
\and J.~Cabrera-Ca\~no		\inst{11}
\and J.~F.~Castander		\inst{12}
\and J.~Cepa			\inst{8,9}
\and M.~Cervi\~no		\inst{2,8}
\and D.~Crist\'obal-Hornillos	\inst{1}
\and A.~Del Olmo		\inst{2}
\and R.~M.~Gonz\'alez Delgado	\inst{2}
\and C.~Husillos		\inst{2}
\and L.~Infante			\inst{13}
\and V.~J.~Mart\'{\i}nez	\inst{6,14}
\and J.~Perea			\inst{2}
\and F.~Prada			\inst{2}
\and J.~M.~Quintana		\inst{2}
}

\institute{Centro de Estudios de F\'{\i}sica del Cosmos de Arag\'on, Plaza San Juan 1, 44001 Teruel, Spain\\\email{clsj@cefca.es} 
\and Instituto de Astrof\'{\i}sica de Andaluc\'{\i}a (IAA-CSIC), Glorieta de la astronom\'{\i}a s/n, 18008 Granada, Spain 
\and Institute for Computational Cosmology, Department of Physics, Durham University, South Road, Durham DH1 3LE, UK 
\and GEPI, Paris Observatory, 77 av. Denfert Rochereau, 75014 Paris, France 
\and Instituto de F\'{\i}sica de Cantabria, Avenida de los Castros s/n, 39005 Santander, Spain 
\and Observatori Astron\`omic, Universitat de Val\`encia, C/ Catedr\'atico Jos\'e Beltr\'an 2, 46980 Paterna, Spain 
\and Observat\'orio Nacional, COAA, Rua General Jos\'e Cristino 77, 20921-400 Rio de Janeiro, Brazil 
\and Instituto de Astrof\'{\i}sica de Canarias, V\'{\i}a L\'actea s/n, La Laguna, 38200 Tenerife, Spain 
\and Departamento de Astrof\'{\i}sica, Facultad de F\'{\i}sica, Universidad de la Laguna, 38200 La Laguna, Spain 
\and Department of Theoretical Physics, University of the Basque Country UPV/EHU, Bilbao, Spain 
\and Departamento de F\'{\i}sica At\'omica, Molecular y Nuclear, Facultad de F\'{\i}sica, Universidad de Sevilla, Spain 
\and Institut de Ci\`encies de l'Espai (ICE-CSIC), Facultat de Ci\'encies, Campus UAB, 08193 Bellaterra, Spain 
\and Departamento de Astronom\'{\i}a, Pontiﬁcia Universidad Cat\'olica. Santiago, Chile 
\and Departament d'Astronomia i Astrof\'{\i}sica, Universitat de Val\`encia, 46100 Burjassot, Spain 
}

\date{Received September 2014}

\abstract
{}
{Our goal is to develop and test a novel methodology to compute accurate close pair fractions with photometric redshifts.}
{We improve the current methodologies to estimate the merger fraction $f_{\rm m}$ from photometric redshifts by (i) using the full probability distribution functions (PDFs) of the sources in redshift space, (ii) including the variation in the luminosity of the sources with $z$ in both the selection of the samples and in the luminosity ratio constrain, and (iii) splitting individual PDFs into red and blue spectral templates to deal robustly with colour selections. We test the performance of our new methodology with the PDFs provided by the ALHAMBRA photometric survey.}
{The merger fractions and rates from the ALHAMBRA survey are in excellent agreement with those from spectroscopic work, both for the general population and for red and blue galaxies. With the merger rate of bright ($M_B \leq -20 - 1.1z$) galaxies evolving as $(1+z)^n$, the power-law index $n$ is larger for blue galaxies ($n = 2.7\pm0.5$) than for red galaxies ($n = 1.3\pm0.4$), confirming previous results. Integrating the merger rate over cosmic time, we find that the average number of mergers per galaxy since $z = 1$ is $N_{\rm m}^{\rm red} = 0.57 \pm 0.05$ for red galaxies and $N_{\rm m}^{\rm blue} = 0.26 \pm 0.02$ for blue galaxies.}
{Our new methodology exploits statistically all the available information provided by photometric redshift codes and provides accurate measurements of the merger fraction by close pairs only using photometric redshifts. Current and future photometric surveys will benefit of this new methodology.}


\titlerunning{The ALHAMBRA survey. Accurate merger fractions from photometric pairs}

\authorrunning{L\'opez-Sanjuan et al.}

\maketitle

\section{Introduction}\label{intro}
In their pioneering study, \citet{toomre72} were able to explain the tails and the distortions of four peculiar galaxies as the intermediate stage of a merger event between two spiral galaxies. Since then, the role of mergers in galaxy evolution has been recognized and studied systematically, both observationally and theoretically. To constraint the role of mergers in galaxy evolution two observational approaches are needed: (i) understand precisely how interactions modify the properties of galaxies and the fate of the merger remnants, and (ii) measure the merger history of different populations over cosmic time to estimate the integrated effect of mergers.

Regarding the first approach, nowadays it is well stated that the major merging (the merger of two galaxies with similar masses, $M_2/M_1 \geq 1/4$) of two spiral galaxies is an efficient mechanism to create new bulge-dominated, red sequence galaxies \citep[][]{naab06ss,rothberg06a,rothberg06b,hopkins08ss,rothberg10,bournaud11}, while major and minor mergers have been proposed as the main mechanism in the mass and size evolution of massive galaxies \citep[e.g.,][]{bezanson09,clsj12sizecos}. In addition, when the separation $r_{\rm p}$ between galaxies in close pairs decreases, the star formation rate (SFR) is enhanced \citep{barton00,lambas03,robaina09,knapen09,patton11} and the metallicity decreases \citep{kewley06,ellison08,scudder12}.

Regarding the second approach, the merger history of a given population is estimated measuring its merger fraction $f_{\rm m}$, i.e., the fraction of galaxies in a sample suffering a merging process, both by morphological criteria (highly distorted galaxies are merger remnants, e.g., \citealt{conselice03,conselice08,cassata05,depropris07,lotz08ff,lotz11,clsj09ffgs,clsj09ffgoods,jogee09,bridge10}), or by close pair statistics (two galaxies close in the sky plane, $r_{\rm p} \leq r_{\rm p}^{\rm max}$, and in redshift space, $\Delta v \leq 500$ km s$^{-1}$, that will lead to a merger, e.g., \citealt{lefevre00,patton00,patton02,patton08,lin04,lin08,depropris05,depropris10,deravel09,deravel11,clsj11mmvvds,clsj13ffmassiv,tasca14}).

Several efforts have been conducted in the literature to study close companions in photometric surveys. Photometric surveys are limited by the $\Delta v$ condition: The $500$ km s$^{-1}$ difference translates to a redshift difference of $|z_1 - z_2| \leq 0.0017(1+z)$, with the best photometric redshifts ($z_{\rm p}$) from current broad+medium-band surveys reaching a precision $\sim0.01(1+z)$ \citep[e.g.,][]{ilbert09,shards,molino13}. Next-generation large photometric redshift surveys will cover huge sky areas ($\gtrsim 5000$ deg$^2$) with broad-band filters, such as the DES (Dark Energy Survey, $grizY$, \citealt{des}) and the LSST (Large Synoptic Survey Telescope, $ugrizY$, \citealt{lsst}), and with narrow-band filters, such as the J-PAS (Javalambre-Physics of the accelerated universe Astrophysical Survey, 56 optical filters of $\sim145\AA$, \citealt{jpas}), providing photometric redshifts for hundreds of million sources. Thus, a suitable and robust methodology to estimate the merger fraction from photometric close pairs is fundamental to exploit the current and the ambitious next photometric surveys.

The most extended approach to tackle with the redshift condition is the estimation of the number of random companions. It can be estimated by either (i) searching for close companions in random positions in the sky, providing the number of expected companions found by chance in a given catalogue \citep[e.g.,][]{kar07,williams11,marmol12,xu12,diazgarcia13,ruiz14}, or (ii) integrating the observed luminosity or mass function over the search area around the central galaxy \citep[e.g.,][]{lefevre00,rawat08,hsieh08,bluck09,bundy09}. Then, the observed number of companions is decontaminated by the random one to obtain the number of real companions.

A probabilistic approach was presented in \citet[][LS10 hereafter]{clsj10pargoods} to deal with the redshift condition. They assume that the probability distribution function (PDF) of the photometric redshifts is well described by a Gaussian. Then, they estimate the overlap between the PDFs of close galaxies in the sky plane to derive the number of pairs per close system. In this approach, each system has a probability of being a real close pair.

\begin{figure}[t]
\centering
\resizebox{\hsize}{!}{\includegraphics{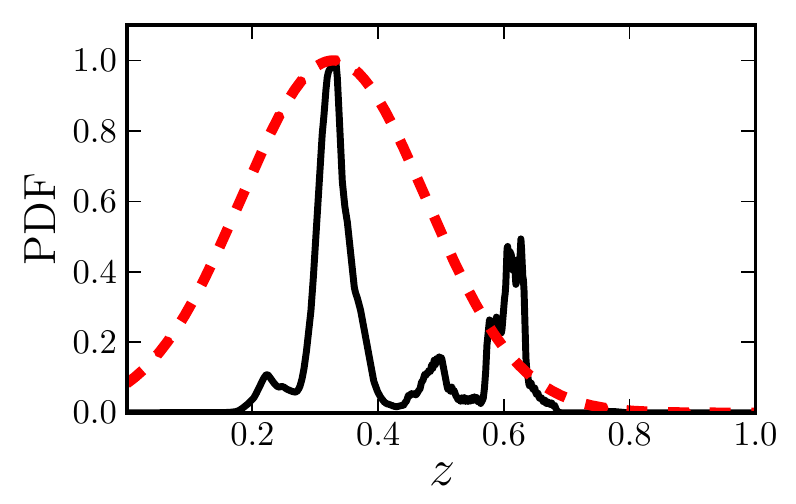}}
\caption{The probability distribution function (PDF) of an ALHAMBRA source (black solid line) with $I = 22.8$ and its Gaussian approach ($z_{\rm p} = 0.329\pm0.147$, red dashed line). Both distributions are normalised to their maximum probability. [{\it A colour version of this plot is available at the electronic edition}].}
\label{PDFgauss}
\end{figure}

However, the previous methods have several shortcomings that should be addressed:

\begin{itemize}
\item The PDFs of those galaxies with low signal-to-noise are poorly approximated by a Gaussian function. We illustrate this point in Fig.~\ref{PDFgauss}. The Gaussian approach of the PDF in this example is $z_{\rm p} = 0.329\pm0.147$, notably worse than the actual PDF that presents two main narrow peaks at $z \sim 0.33$ and $z \sim 0.61$. We note that several studies have proven that the PDFs are the best approach to deal with photometric redshifts \citep[e.g.,][]{soto02,cunha09,wittman09,myers09,schmidt13,carrasco14a}.
\item The luminosity, the stellar mass, or the star-formation rate of a source depend on its redshift. Even in the Gaussian approach, previous studies assume the properties of galaxies as constant with redshift, and they set them to the values of the best photometric redshift solution. This is a crude approximation that impacts the selection of the samples, as well as the luminosity and the mass difference between the galaxies in pairs. 
\item The colour selection is blurred by photometric errors. As blue galaxies spill over the red locus and vice versa, the differences between both populations diminish for the galaxies with low signal-to-noise. 
\end{itemize}

In the present paper we solve the previous shortcomings by generalising and extending the LS10 methodology. The new method (i) uses the full PDFs of the sources in redshift space, (ii) includes the variation in the luminosity of the sources with $z$ in both the selection of the samples and the luminosity ratio constrain, and (iii) splits individual PDFs into red and blue spectral templates to deal robustly with colour selections. We take advantage of the unique design, depth, and photometric redshift accuracy of the ALHAMBRA\footnote{http://alhambrasurvey.com} (Advanced, Large, Homogeneous Area, Medium-Band Redshift Astronomical) photometric survey \citep{alhambra} to develop and test our new methodology. 

The paper is organised as follows. In Sect.~\ref{data} we present the ALHAMBRA survey and its photometric redshifts. We develop the methodology to measure accurate merger fractions by PDF analysis of photometric close pairs in Sect.~\ref{metodo}. We test our new methodology by comparison with spectroscopic studies in Sects.~\ref{ffsec} and \ref{RMMsec}, and in Sect.~\ref{conclusions} we summarise our work and present our conclusions. Throughout this paper we use a standard cosmology with $\Omega_{\rm m} = 0.307$, $\Omega_{\Lambda} = 0.693$, $H_{0}= 100h$ km s$^{-1}$ Mpc$^{-1}$, and $h = 0.678$ \citep{planck13}. Magnitudes are given in the AB system \citep{oke83}.

\begin{table*}
\caption{The ALHAMBRA survey fields.}
\label{alhambra_fields_tab}
\begin{center}
\begin{tabular}{lcccc}
\hline\hline\noalign{\smallskip}
Field      &    Overlapping     &    RA    &    DEC     &    sub-fields / area \\
name       &      survey	&   (J2000) & (J2000)    &   (\# / deg$^2$)\\
\noalign{\smallskip}
\hline
\noalign{\smallskip}
ALHAMBRA-2  &  DEEP2 	\citep{deep2}	& 01 30 16.0	& +04 15 40   &  8 / 0.377	\\
ALHAMBRA-3  &  SDSS  	\citep{sdssdr8}	& 09 16 20.0	& +46 02 20   &  8 / 0.404	\\
ALHAMBRA-4  &  COSMOS  	\citep{cosmos}  & 10 00 00.0	& +02 05 11   &  4 / 0.203	\\
ALHAMBRA-5  &  GOODS-N	\citep{goods}	& 12 35 00.0	& +61 57 00   &  4 / 0.216	\\
ALHAMBRA-6  &  AEGIS	\citep{aegis}	& 14 16 38.0	& +52 24 50   &  8 / 0.400	\\
ALHAMBRA-7  &  ELAIS-N1	\citep{elais}	& 16 12 10.0	& +54 30 15   &  8 / 0.406	\\
ALHAMBRA-8  &  SDSS	\citep{sdssdr8}	& 23 45 50.0	& +15 35 05   &  8 / 0.375	\\
Total	    &		&		&	      & 48 / 2.381\\
\noalign{\smallskip}
\hline
\end{tabular}
\end{center}
\end{table*}

\section{The ALHAMBRA survey}\label{data}

The ALHAMBRA survey provides a photometric data set over 20 contiguous, equal-width ($\sim$300\AA), non-overlapping, medium-band optical filters (3500\AA -- 9700\AA) plus 3 standard broad-band near-infrared (NIR) filters ($J$, $H$, and $K_{\rm s}$) over 8 different regions of the northern sky \citep{alhambra}. The survey has the aim of understanding the evolution of galaxies along cosmic time by sampling a large enough cosmological fraction of the Universe, for which reliable spectral energy distributions (SEDs) and precise photometric redshifts are needed. The simulations of \citet{benitez09}, which relate the image depth and the accuracy of the photometric redshifts to the number of filters, suggested that the filter set chosen for ALHAMBRA can achieve a photometric redshift precision that is three times better than a classical $4 - 5$ optical broad-band filter set. This expectation is confirmed by the results presented in \citet{molino13}. The final survey parameters and scientific goals, as well as the technical properties of the filter set, were described by \citet{alhambra}. The survey has collected its data for the 20+3 optical-NIR filters with the 3.5m telescope at the Calar Alto observatory, using the wide-field camera LAICA (Large Area Imager for Calar Alto) in the optical and the OMEGA–2000 camera in the NIR. The full characterisation, description, and performance of the ALHAMBRA optical photometric system were presented in \citet{aparicio10}. A summary of the optical reduction can be found in Crist\'obal-Hornillos et al. (in prep.), whereas the NIR reduction is in \citet{cristobal09}.

The ALHAMBRA survey has observed eight well-separated regions of the northern sky. The wide-field camera LAICA has four chips with a $15\arcmin \times 15\arcmin$ field-of-view per chip (0.22 arcsec/pixel). The separation between chips is also $15\arcmin$. Thus, each LAICA pointing provides four separated areas in the sky. Currently, six ALHAMBRA regions comprise two LAICA pointings. In these cases, the pointings define two separate strips in the sky. In our study, we assumed the four chips in each pointing as independent sub-fields. The photometric calibration of the field ALHAMBRA-1 is currently ongoing, and the fields ALHAMBRA-4 and ALHAMBRA-5 comprise one pointing each \citep[see][for details]{molino13}. We summarise the properties of the seven ALHAMBRA fields used in the present paper in Table~\ref{alhambra_fields_tab}. At the end, the data we used comprise 48 sub-fields of $\sim180$ arcmin$^2$ each, which can be assumed as independent for merger fraction studies as demonstrated by \citet{clsj14ffcosvar}. 

\subsection{Bayesian photometric redshifts in ALHAMBRA}
We rely on the ALHAMBRA photometric redshifts to compute the merger fraction. The photometric redshifts used all over the present paper are fully presented and tested in \citet{molino13}, and we summarise their principal characteristics below.

The photometric redshifts of ALHAMBRA were estimated with BPZ2.0, a new version of BPZ \citep[Bayesian Photometric Redshift,][]{benitez00} estimator. BPZ is a SED-fitting method based in a Bayesian inference, where a maximum likelihood is weighted by a prior probability. The library of 11 SEDs, that comprises 4 ellipticals (E), 1 lenticular (S0), 2 spirals (S), and 4 starbursts (SB), and the prior probabilities used by BPZ2.0 in ALHAMBRA are detailed in Ben\'{\i}tez (in prep.). ALHAMBRA relied on the ColorPro software \citep{colorpro} to perform PSF-matched aperture-corrected photometry, which provided both total magnitudes and isophotal colours for the galaxies. In addition, an homogeneous photometric zero point recalibration was done using either spectroscopic redshifts (when available) or accurate photometric redshifts from emission-line galaxies \citep{molino13}. Sources were detected in a synthetic $F814W$ filter image, as noted $I$ in the following, defined to resemble the HST/$F814W$ filter. The areas of the images affected by bright stars, as well as those with lower exposure times (e.g., the edges of the images), were masked following \citet{arnaltemur14}. The total area covered by the current ALHAMBRA data after masking is 2.38 deg$^{2}$ (Table~\ref{alhambra_fields_tab}). Finally, a statistical star/galaxy separation is encoded in the variable \texttt{Stellar\_Flag} of the ALHAMBRA catalogues, and throughout this paper, we keep those ALHAMBRA sources with $\texttt{Stellar\_Flag} \leq 0.5$ as galaxies.

The photometric redshift accuracy, as estimated by comparison with $\sim 7200$ spectroscopic redshifts ($z_{\rm s}$), is encoded in the normalized median absolute deviation (NMAD) of the photometric versus spectroscopic redshift distribution \citep{ilbert06,eazy},
\begin{equation}
\sigma_{\rm NMAD} = 1.48 \times {\rm median}\,\bigg( \frac{|\,\delta_z - {\rm median}(\delta_z)\,|}{1 + z_{\rm s}} \bigg),
\end{equation} 
where $\delta_z = z_{\rm p} - z_{\rm s}$. The fraction of catastrophic outliers $\eta$ is defined as the fraction of galaxies with $|\,\delta_z\,|/(1 + z_{\rm s}) > 0.2$. In the case of ALHAMBRA, $\sigma_{\rm NMAD} = 0.011$ for $I \leq 22.5$ galaxies with a fraction of catastrophic outliers of $\eta = 2.1$\%. We refer to \citet{molino13} for a more detailed discussion.

The \texttt{odds} quality parameter, as noted $\mathcal{O}$, is a proxy for the photometric redshift reliability of the sources and is also provided by BPZ2.0. The $\mathcal{O}$ parameter is defined as the redshift probability enclosed on a $\pm K(1+z)$ region around the main peak in the PDF of the source, where the constant $K$ is specific for each photometric survey. \citet{molino13} find that $K = 0.0125$ is the optimal value for ALHAMBRA since this is the expected averaged accuracy for most galaxies in the survey. Thus, $\mathcal{O} \in [0,1]$ and it is related to the confidence of the photometric redshifts, making it possible to derive high quality samples with better accuracy and lower rate of catastrophic outliers. For example, a $\mathcal{O} \geq 0.5$ selection for $I \leq 22.5$ galaxies yields $\sigma_{\rm NMAD} = 0.009$ and $\eta = 1$\%, while $\sigma_{\rm NMAD} = 0.006$ and $\eta = 0.8$\% if galaxies with $\mathcal{O} \geq 0.9$ are selected \citep[see][for further details]{molino13}. \citet{clsj14ffcosvar} set $\mathcal{O} \geq 0.3$ as the optimal selection for merger fraction studies in ALHAMBRA. We study the impact of this $\mathcal{O}$ selection in Sect.~\ref{osr}.

\begin{figure}[t]
\centering
\resizebox{\hsize}{!}{\includegraphics{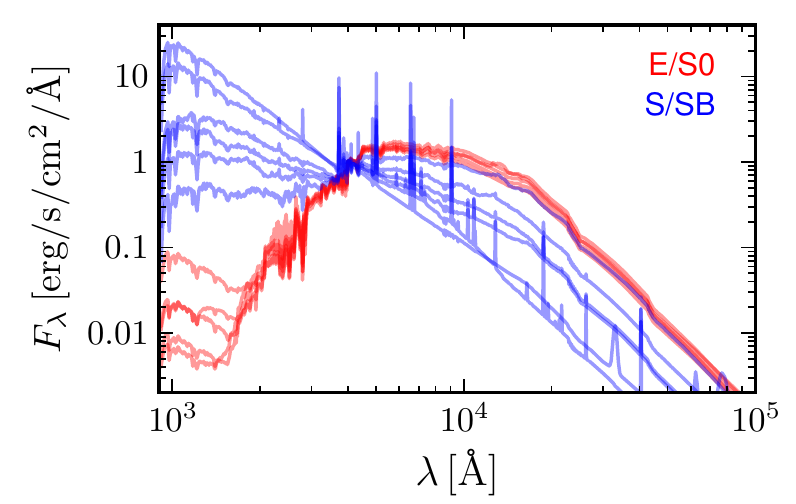}}
\caption{Spectral energy distributions of the red ($T = {\rm E/S0}$) and blue ($T = {\rm S/SB}$) templates in BPZ2.0. The templates are normalised at $\lambda = 4000\AA$ for clarity. [{\it A colour version of this plot is available at the electronic edition}].}
\label{Tredblue}
\end{figure}

\begin{figure}[t]
\centering
\resizebox{\hsize}{!}{\includegraphics{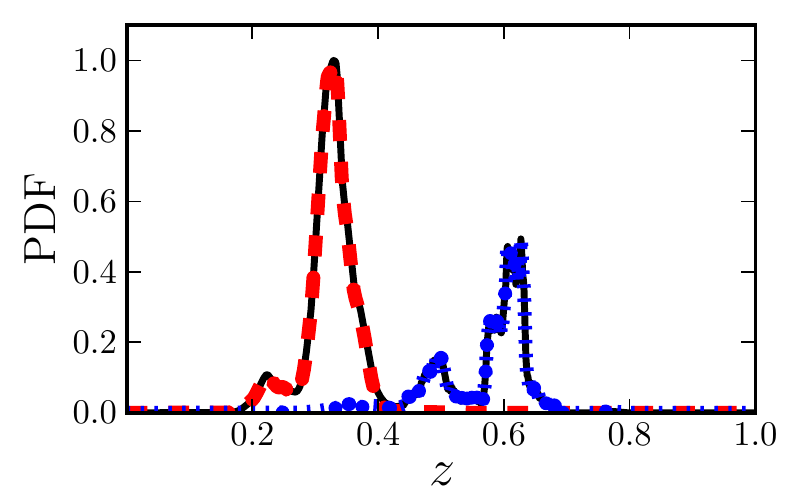}}
\resizebox{\hsize}{!}{\includegraphics{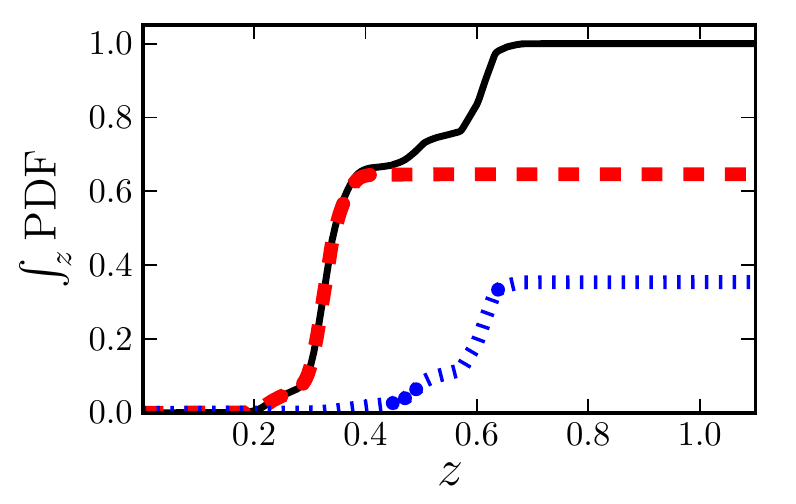}}
\caption{Partial probability distribution functions ({\it top panel}) and the cumulative distribution functions ({\it bottom panel}) of the source presented in Fig.~\ref{PDFgauss}. The black solid lines mark the total PDF, the red dashed lines mark the red templates, ${\rm PDF}^{\rm red} = {\rm PDF}\,(z,{\rm E/S0})$, and the blue dotted lines mark the blue templates, ${\rm PDF}^{\rm blue} = {\rm PDF}\,(z,{\rm S/SB})$. This galaxy counts as 0.65 red and 0.35 blue in the analysis ({\it bottom panel}). [{\it A colour version of this plot is available at the electronic edition}].}
\label{PDFredblue}
\end{figure}

\subsection{Probability distribution functions in ALHAMBRA}\label{pdfs}
This section is devoted to the description of the probability distribution functions of the ALHAMBRA sources. The probability of a galaxy $i$ being located at redshift $z$ and having a spectral type $T$ is $\mathrm{PDF}_{i}\,(z,T)$. This probability function is the posterior provided by BPZ2.0. The probability of the galaxy $i$ of being located at redshift $z$ is then (Fig.~\ref{PDFgauss})
\begin{equation}
\mathrm{PDF}_{i}\,(z) = \int \mathrm{PDF}_{i}\,(z,T)\,{\rm d}T.
\end{equation}
Moreover, the total probability of the galaxy $i$ of being located at $z_1 \leq z \leq z_2$ is
\begin{equation}
P_{i}\,(z_1,z_2) = \int_{z_1}^{z_2} \mathrm{PDF}_{i}\,(z)\,{\rm d}z.
\end{equation}
The distribution function $\mathrm{PDF}\,(z,T)$ is normalised to one by definition, this is, there is {\it one galaxy} spread over the redshift and template spaces. Formally, 
\begin{equation}
1 = \int \mathrm{PDF}_{i}\,(z)\,{\rm d}z  = \int\!\!\!\int \mathrm{PDF}_{i}\,(z,T)\,{\rm d}T\,{\rm d}z.
\end{equation}

In the present paper, the definition of red and blue galaxies takes advantage of the profuse information encoded in the PDFs. Instead of selecting galaxies according to their observed colour or their best spectral template, we split each PDF into ``red'' templates ($T = {\rm E/S0}$), as noted ${\rm PDF}^{\rm red}$, and ``blue'' templates ($T = {\rm S/SB}$), as noted ${\rm PDF}^{\rm blue}$ (Fig.~\ref{Tredblue}). This is, a given galaxy can be both red and blue (Fig~\ref{PDFredblue}). Formally,
\begin{eqnarray}
\mathrm{PDF}_{i}\,(z) &=& \mathrm{PDF}^{\rm red}_{i}(z) + \mathrm{PDF}^{\rm blue}_{i}(z)\nonumber\\ &=& \int \mathrm{PDF}_{i}\,(z,{\rm E/S0})\,{\rm d}T + \int \mathrm{PDF}_{i}\,(z,{\rm S/SB})\,{\rm d}T.
\end{eqnarray}
In practice, the red templates have $T \in [1,5.5]$ and the blue templates have $T \in (5.5,11]$ in the ALHAMBRA catalogues. This is a major step forward in the methodology, that is able to robustly deal with colour segregations without any pre-selection of the sources.

The $B-$band absolute magnitude of a galaxy with observed magnitude $I = 20$, spectral type $T$, and located  at redshift $z$ is noted as $M_B^{20}\,(z,T)$, which is also provided by BPZ2.0. In the present paper, we are interested in $M_B$ as a function of $z$. We estimate $M_B\,(z)$ as
\begin{equation}
M_B\,(z) = \frac{\int M_B^{20}\,(z,T) \times {\rm PDF}(z,T)\,{\rm d}T}{\int {\rm PDF}(z,T)\,{\rm d}T} + (I - 20).\label{mbz}
\end{equation}
The average $B-$band absolute magnitude of a galaxy is then
\begin{equation}
\langle M_B \rangle = \frac{\int M_B\,(z) \times {\rm PDF}(z)\,{\rm d}z}{\int {\rm PDF}(z)\,{\rm d}z}.
\end{equation}

We will now see how, thanks to the probability functions defined in this section, we are able to statistically use the output of current photometric redshift codes without losing information. This is capital to perform accurate and robust studies of the merger fraction and the environment with photometric redshifts.

\subsection{Sample selection}\label{selection}
Throughout this paper, we focus our analysis on the galaxies in the ALHAMBRA first data release\footnote{http://cloud.iaa.es/alhambra/}. This catalogue comprises $\sim500$k sources and is complete ($5\sigma$, $3\arcsec$ aperture) for $I \leq 24.5$ galaxies \citep{molino13}.

We perform our study in a given redshift range $z \in [z_{\rm min}, z_{\rm max})$, and in samples selected with $B-$band luminosity. To define the galaxy samples under study, we first estimated the $B-$band selection function, $\mathcal{S}\,(z)$, as
\begin{equation}
\mathcal{S}\,(z) = \left\{\begin{array}{ll}
1, & \quad {\rm if}\ M_{B}^{\rm bri} < M_{B}\,(z) + Qz \leq M_{B}^{\rm sel}, \\
0, & \quad {\rm otherwise},
\end{array}\right.
\label{MBmask}
\end{equation}
where $M_{B}(z)$ is the $B-$band luminosity of the galaxy from Eq.~(\ref{mbz}), the term $Qz$ accounts for the evolution of the luminosity function with redshift \citep[e.g.,][]{lin08}, $M_{B}^{\rm sel}$ is the selection magnitude of the sample, and $M_{B}^{\rm bri}$ imposes a maximum luminosity in the study. We assume $M_{B}^{\rm bri} = -22$ in this paper to avoid the different clustering properties of the brightest galaxies \citep{patton00,patton02,lin08}. Then, we kept as galaxies in the sample those sources with
\begin{equation}
\int_{z_{\rm min}}^{z_{\rm max}} {\rm PDF}_i\,(z) \times \mathcal{S}^i\,(z)\,{\rm d}z > 0.
\label{MBsel}
\end{equation}
We note that Eq.~(\ref{MBmask}) defines a $B-$band luminosity selection, but the selection function can be defined in the same way for mass-selected samples if the stellar mass $M_{\star}\,(z)$ of the sources is known.

\section{Measuring of the merger fraction in photometric samples by PDF analysis}\label{metodo}

In this section, we recall first the methodology to compute the merger fraction from spectroscopic close pairs (Sect.~\ref{ffspec}), and then we develop the extension of the method to the photometric redshift regime (Sects.~\ref{ffphot} and \ref{ffpdf}). The statistical weights devoted to correcting for the selection effects in the photometric case are defined in Sect.~\ref{weights}. Finally, the output of the code is detailed in Sect.~\ref{output}.

\subsection{The merger fraction in spectroscopic samples}\label{ffspec}
The linear distance between two sources can be obtained from their projected separation, $r_{\rm p} = \theta\,d_A(z_1)$, and their rest-frame relative velocity along the line of sight, $\Delta v = {c\, |z_2 - z_1|}/(1+z_1)$, where $z_1$ and $z_2$ are the redshift of the central (the most luminous galaxy in the pair) and the satellite galaxy, respectively; $\theta$ is the angular separation, in arcsec, of the two galaxies on the sky plane; and $d_A(z)$ is the angular diameter distance, in kpc arcsec$^{-1}$, at redshift $z$. Two galaxies are defined as a close pair if $r_{\rm p}^{\rm min} \leq r_{\rm p} \leq r_{\rm p}^{\rm max}$ and $\Delta v \leq \Delta v^{\rm max}$. To ensure well de-blended sources and to minimise colour contamination in ground-based surveys, the minimum search radius is usually $r_{\rm p}^{\rm min} \geq 5h^{-1}$ kpc. With $r_{\rm p}^{\rm max} \leq 100h^{-1}$ kpc and $\Delta v^{\rm max} \leq 500$ km s$^{-1}$, 50\% to 70\% of the selected close pairs will finally merge \citep{patton00,patton08,bell06,jian12}.

To compute the merger fraction, one defines a primary and a secondary sample. The primary sample comprises the population of interest and one looks for those galaxies in the secondary sample that fulfil the close pair criterion for each galaxy of the primary sample. With the previous definitions the merger fraction is
\begin{equation}
f_{\rm m}\ = \frac{N_{\rm p}}{N_{1}},\label{ncspec}
\end{equation}
where $N_1$ is the number of sources in the primary sample and $N_{\rm p}$ the number of close pairs. This definition applies to spectroscopic volume-limited samples, but we rely on photometric redshifts to compute $f_{\rm m}$. In the following, we expand the methodology presented by LS10 to use an arbitrary PDF in redshift space and to take into account the variation of galaxy properties with $z$. 

\subsection{PDF analysis of photometric close pairs}\label{ffphot}
In this section, we detail the steps in the computation of the merger fraction in photometric redshift surveys. The primary and the secondary samples were defined thanks to the $B-$band selection function introduced in Sect.~\ref{selection}, as noted $\mathcal{S}_{1}\,(z)$ for the primary sample and $\mathcal{S}_{2}\,(z)$ for the secondary sample.

\subsubsection{Initial list of projected companions}\label{ppf:0}
To define the initial list of projected close companions, we estimated the maximum angular separation possible in the first instance, as noted $\theta_{\rm top}$. This angular separation is defined as 
\begin{equation}
\theta_{\rm top} = \frac{r_{\rm p}^{\rm max}}{d_{\rm A}(z_{\rm min})}.
\end{equation}
Then, for each galaxy in the primary sample, we searched those galaxies in the secondary sample with $\theta \leq \theta_{\rm top}$. We end this first step with a list of systems composed by a principal source and its projected companions.

\begin{figure}[t]
\centering
\resizebox{\hsize}{!}{\includegraphics{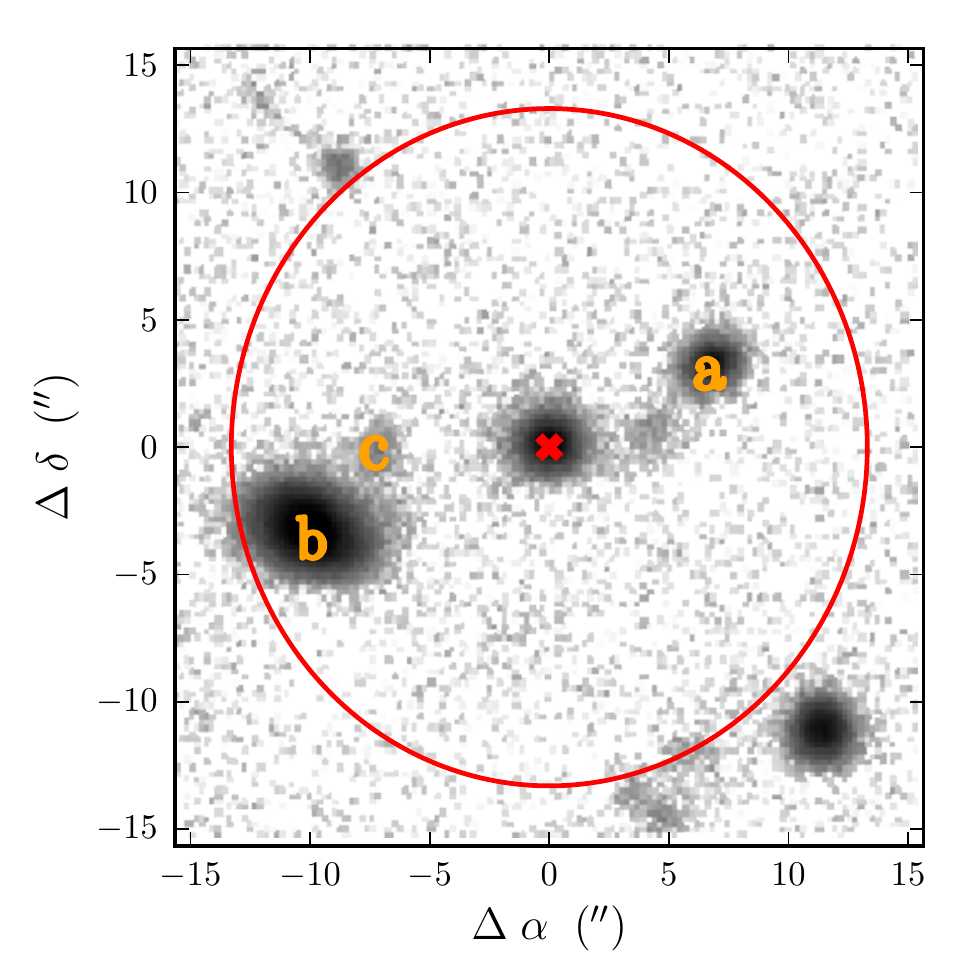}}
\caption{Postage stamp of a particular ALHAMBRA system composed by a principal source (red cross) and its three projected companions in the sky plane (orange letters) in the $I$ band. North is up and East is left. The axes show the right ascension ($\alpha$) and the declination ($\delta$) offset with respect to the position of the principal source ($\alpha_1 = 188.7021$, $\delta_1 = 61.9441$). The red circle marks the angular separation in the sky plane for $r_{\rm p}^{\rm max} = 50h^{-1}$ kpc at $z_{\rm min} = 0.4$, $\theta_{\rm top} = 13.32\arcsec$.  [{\it A colour version of this plot is available at the electronic edition}].}
\label{ppf:0:fig}
\end{figure}

To illustrate the performance of our method and for the sake of clarity, we present a particular ALHAMBRA system as an example in the following. We defined the primary sample with absolute $B-$band magnitude $M_{B,1}^{\rm sel} = -20$ and the secondary sample with $M_{B,2}^{\rm sel} = -18.5$, and assumed an evolution in the selection of $Q = 1.1$. We used $r_{\rm p}^{\rm min} = 10h^{-1}$ kpc as minimum search radius, $r_{\rm p}^{\rm max} = 50h^{-1}$ kpc as maximum search radius, $z_{\rm min} = 0.4$ as the minimum redshift in the study, and $z_{\rm max} = 1$ as the maximum redshift in the study. The parameters above are similar to those used in Sect.~\ref{RMMsec}. The principal galaxy of the ``system zero'' is located at $\alpha_1 = 188.7021$ and $\delta_1 = 61.9441$. We found $\theta_{\rm top} = 13.32\arcsec$ with the assumed parameters. We searched companions in the secondary sample, and we found three projected companions (Fig.~\ref{ppf:0:fig}). We note that the companions {\it a} and {\it b} are also in the primary sample.

\begin{figure}[t]
\centering
\resizebox{\hsize}{!}{\includegraphics{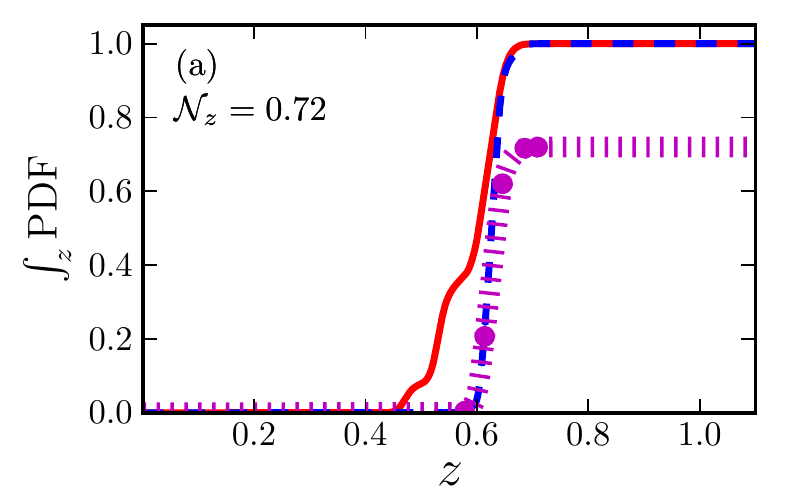}}
\resizebox{\hsize}{!}{\includegraphics{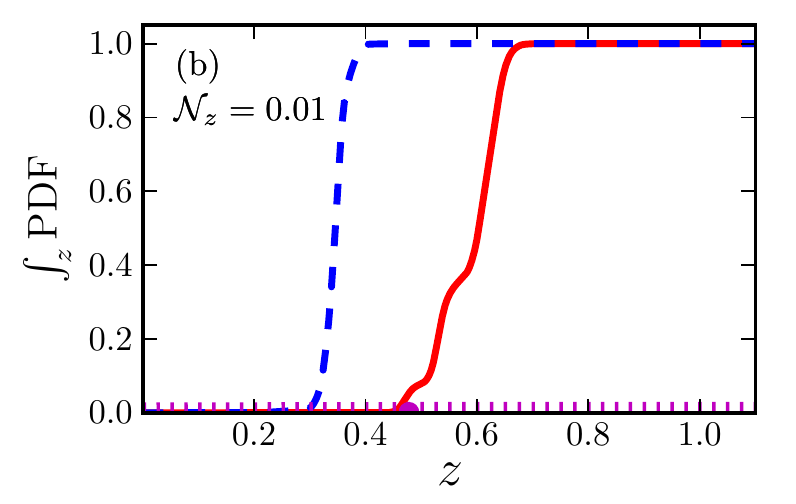}}
\resizebox{\hsize}{!}{\includegraphics{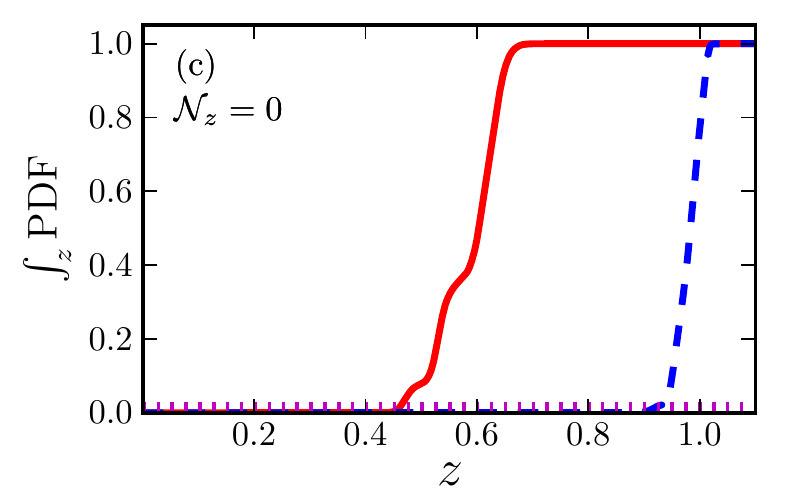}}
\caption{Cumulative distribution function of the principal galaxy (red solid line), its projected companions (blue dashed lines), and the $\mathcal{Z}$ function of the close pairs (purple dotted lines). The letter in each panel refers to the companion galaxy in Fig.~\ref{ppf:0:fig}. The number of pairs in each system, $\mathcal{N}_z$, is labelled in the panels. [{\it A colour version of this plot is available at the electronic edition}].}
\label{ppf:1:fig}
\end{figure}

\subsubsection{The redshift probability $\mathcal{Z}$}\label{ppf:1}
In the initial list defined above, a galaxy can have more than one projected companion. In that case, we took each possible pair separately, i.e., if the companion galaxies {\it a}, {\it b} and {\it c} are close to the principal galaxy {\it X}  (Fig.~\ref{ppf:0:fig}), we study the central--satellite pairs {\it X}--{\it a}, {\it X}--{\it b}, and {\it X}--{\it c} independently. This defines the initial list of projected close pairs. We note that if the galaxies {\it X} and {\it a} are both in the primary sample, the close pairs {\it X}--{\it a} and {\it a}--{\it X} could be present in the initial list. We cleaned the initial list for duplicates before starting the study in the redshift space, keeping the galaxy with lower $\langle M_B \rangle$ as the central galaxy in the pair.

For each projected close pair in the initial list, we define the redshift probability function $\mathcal{Z}$ as 
\begin{equation}
\mathcal{Z}(z) = \frac{2 \times {\rm PDF}_1 (z) \times {\rm PDF}_2 (z)}{{\rm PDF}_1 (z) + {\rm PDF}_2 (z)} = \frac{{\rm PDF}_1 (z) \times {\rm PDF}_2 (z)}{N(z)},\label{nuk}
\end{equation}
where
\begin{equation}
N(z) = \frac{{\rm PDF}_1 (z) + {\rm PDF}_2 (z)}{2}.
\end{equation}

We convolve the PDFs of the central galaxy (${\rm PDF}_1$) and its satellite (${\rm PDF}_2$) to obtain the shape of the function $\mathcal{Z}$, and we normalise to the number of potential pairs (2 galaxies per pair) at each redshift, as noted $N(z)$. We note that $\int N(z)\,{\rm d}z = 1$ by construction. This normalisation is capital in the methodology because it brings the close pair systems to a common scale, with $\mathcal{Z}(z)$ being the {\it number of close pairs in the system at redshift $z$}. Thus, the integral of the function $\mathcal{Z}$ provides the total number of pairs in the system, $\mathcal{N}_z$. We only kept those projected close pairs with $\mathcal{N}_z = \int \mathcal{Z}(z)\,{\rm d}z > 0$ in the subsequent analysis.

The cumulative PDFs and the derived $\mathcal{Z}$ functions for the three projected close pairs in the ``system zero'' are shown in Fig.~\ref{ppf:1:fig}. The PDF of the central galaxy is the same in all the panels, with the best photometric redshift $z_{\rm p,1} = 0.604$. The first companion, {\it panel (a)}, has a photometric redshift $z_{\rm p,2} = 0.630$ and the overlap of the PDFs is evident, with $\mathcal{N}_z = 0.72$. The second companion, {\it panel (b)}, has $z_{\rm p,2} = 0.350$ and the PDFs overlap marginally, with $\mathcal{N}_z = 0.01$. Despite the low probability of this close pair, we kept it in the subsequent analysis because $\mathcal{N}_z > 0$. The third companion, {\it panel (c)}, has $z_{\rm p,2} = 0.988$ and the PDFs do not overlap, with $\mathcal{N}_z = 0$. Thus, we discard this close pair in the following.

\subsubsection{The angular mask $\mathcal{M}^{\theta}$}\label{ppf:2}

\begin{figure}[t]
\centering
\resizebox{\hsize}{!}{\includegraphics{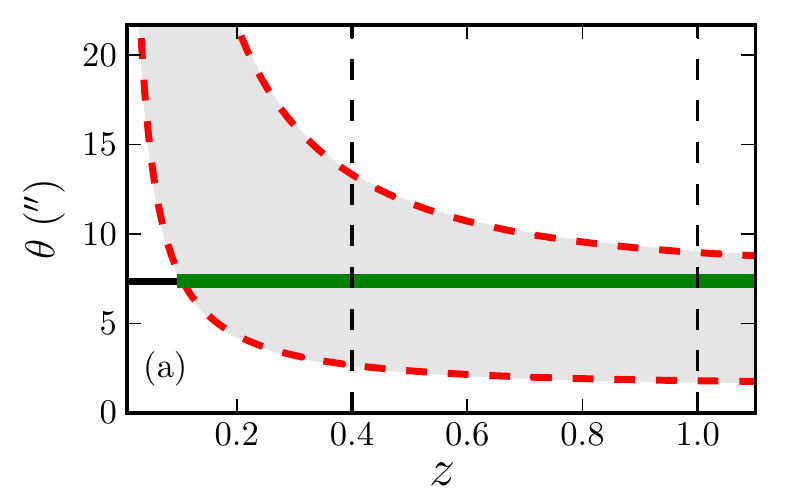}}
\resizebox{\hsize}{!}{\includegraphics{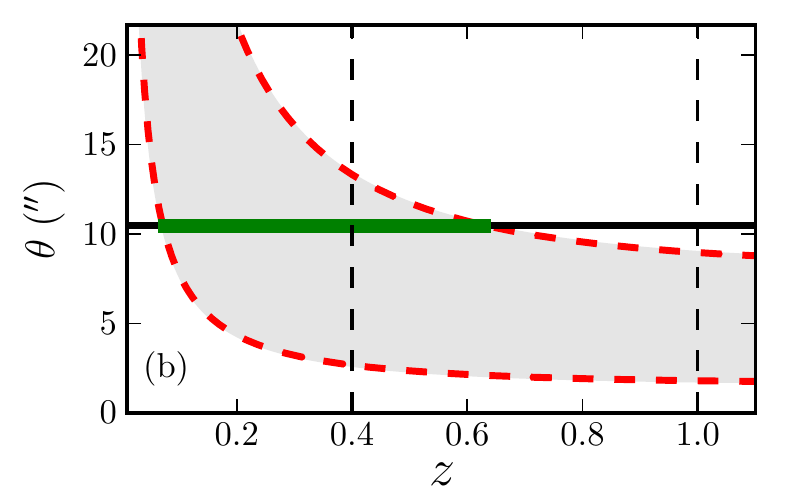}}
\caption{Angular separation $\theta$ as a function of redshift. The solid lines mark the measured angular separation of the close pairs, $\theta = 7.36\arcsec$ in {\it panel (a)} and $\theta = 10.46\arcsec$ in {\it panel (b)}. The letter in each panel refers to the companion galaxy in Fig.~\ref{ppf:0:fig}. The grey area marks the angular separations between $\theta_{\rm min}\,(z)$ and $\theta_{\rm max}\,(z)$ (red dashed lines). The vertical dashed lines mark the redshift range under study, $0.4 < z < 1$. The redshifts at which the angular mask $\mathcal{M}^{\theta}$ is equal to one are marked with the thick green line. [{\it A colour version of this plot is available at the electronic edition}].}
\label{ppf:2:fig}
\end{figure}

The function $\mathcal{Z}$ defined in the previous section only accounts for the overlap of the central and the satellite galaxy probabilities in redshift space. However, the definition of a close pair also includes conditions on the projected distance $r_{\rm p}$ and on the luminosity of the sources. Thus, the next step was to define {\it redshift masks}, as noted $\mathcal{M}\,(z)$, to account for the other conditions of interest. These masks complement the function $\mathcal{Z}$ and they are equal to one at those redshifts where a particular condition is fulfilled, and equal to zero otherwise. 

The first mask that we computed is the {\it angular mask} $\mathcal{M}^{\theta}$. The function $d_A$ changes with redshift, so a close pair in the sky plane at $z_{\rm min}$ (Sect.~\ref{ppf:0}) might not be a close pair at higher redshifts. Thus, we estimated the functions $\theta_{\rm min}\,(z) = r_{\rm p}^{\rm min} / d_A\,(z)$ and $\theta_{\rm max}\,(z) = r_{\rm p}^{\rm max} / d_A\,(z)$, and imposed the condition $\theta_{\rm min}\,(z) \leq \theta \leq \theta_{\rm max}\,(z)$. Formally,
\begin{equation}
\mathcal{M}^{\theta}(z) = \left\{\begin{array}{ll}
1, & \quad {\rm if}\ \theta_{\rm min}\,(z) \leq \theta \leq \theta_{\rm max}\,(z), \\
0, & \quad {\rm otherwise}.
\end{array}\right.
\end{equation}

The measured angular separation of the close pairs in the ``system zero'' are shown in Fig.~\ref{ppf:2:fig}. The first pair, {\it panel (a)}, has $\theta = 7.36\arcsec$ and fulfils the angular condition at $z > 0.106$. The second pair, {\it panel (b)}, has $\theta = 10.46\arcsec$ and fulfils the angular condition in two redshift ranges, $0.071 < z < 0.663$ and $z > 4.099$ (outside the plotted redshift range). We note that the present paper is focused at $z < 1$, but the method has been developed to study the close pairs in the full redshift space.

\subsubsection{The pair selection mask $\mathcal{M}^{\rm pair}$}\label{ppf:3}

\begin{figure}[t]
\centering
\resizebox{\hsize}{!}{\includegraphics{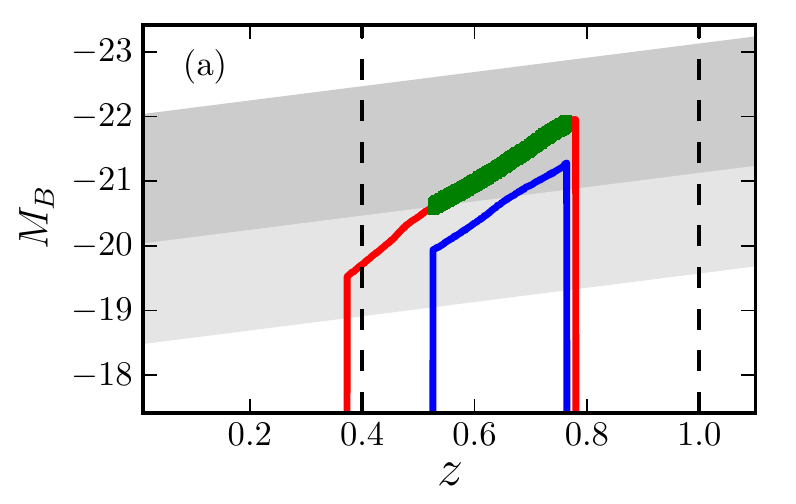}}
\resizebox{\hsize}{!}{\includegraphics{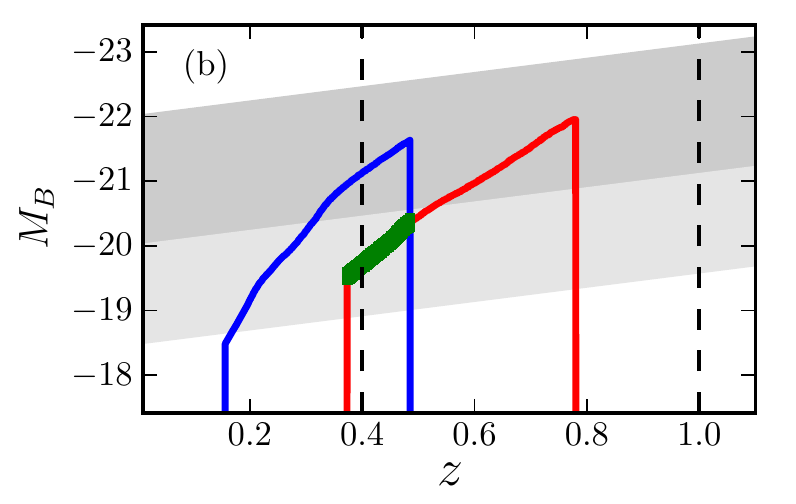}}
\caption{$B-$band absolute magnitude $M_B$ as a function of redshift. The red and blue solid lines mark the absolute magnitude of the principal and the companion galaxy, respectively, at those redshifts with ${\rm PDF}\,(z) > 0$. The letter in each panel refers to the companion galaxy in Fig.~\ref{ppf:0:fig}. The dark grey area marks the selection of the primary sample, and both grey areas mark the selection of the secondary sample (i.e., the primary sample is included in the secondary one). The vertical dashed lines mark the redshift range under study, $0.4 < z < 1$. The redshifts at which the pair selection mask $\mathcal{M}^{\rm pair}$ is equal to one are marked with the thick green line. [{\it A colour version of this plot is available at the electronic edition}].}
\label{ppf:3mb:fig}
\end{figure}

\begin{figure}[t]
\centering
\resizebox{\hsize}{!}{\includegraphics{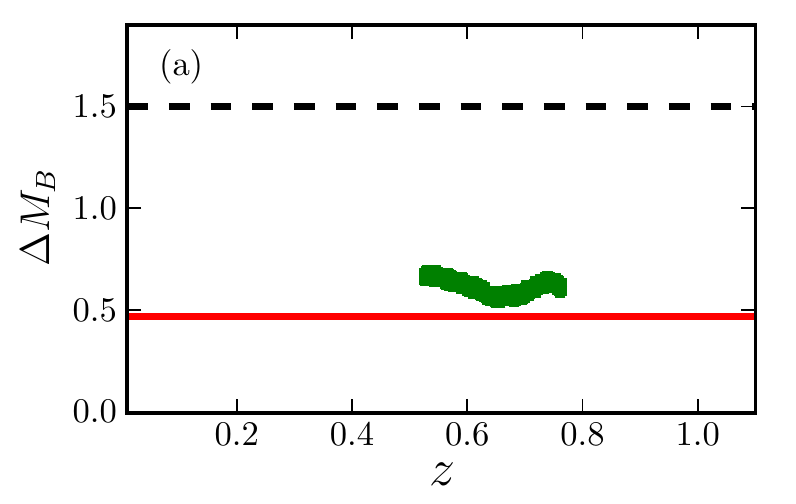}}
\resizebox{\hsize}{!}{\includegraphics{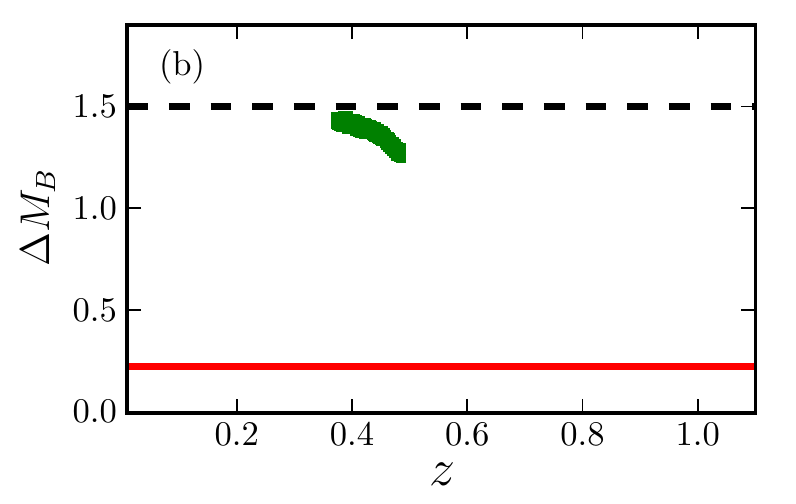}}
\caption{Absolute magnitude difference between the galaxies in the pair, $\Delta M_{B}\,(z) = |M_{B,2}(z) - M_{B,1}(z)|$, as a function of redshift (thick green line).  The letter in each panel refers to the companion galaxy in Fig.~\ref{ppf:0:fig}. The dashed lines mark the major merger limit, $\Delta M_{B} = 1.5$. The red solid lines mark the $\Delta M_{B}$ computed with the absolute magnitudes estimated at the best photometric redshifts. [{\it A colour version of this plot is available at the electronic edition}].}
\label{ppf:3Delta:fig}
\end{figure}

In this section we define the pair selection mask, as noted $\mathcal{M}^{\rm pair}(z)$. The pair selection mask imposes three conditions simultaneously: the selection of the primary sample, the selection of the companion sample, and the luminosity ratio between the galaxies in the pair. The last condition is needed to define major and minor companions. Formally, the general form of the pair selection mask is
\begin{equation}
\mathcal{M}^{\rm pair}(z) = \left\{\begin{array}{ll}
1, & \quad {\rm if}\ M_{B}^{\rm bri} < M_{B,1}(z) + Qz \leq M_{B,1}^{\rm sel}, \\
   & \quad \quad M_{B}^{\rm bri} < M_{B,2}(z) + Qz \leq M_{B,2}^{\rm sel}, \\
   & \quad \quad \Delta M_{B}\,(z) \leq -2.5\log_{10}\mu,\\
0, & \quad {\rm otherwise},
\end{array}\right.
\label{Mselmask}
\end{equation}
where $\Delta M_{B}\,(z) = |M_{B,2}(z) - M_{B,1}(z)|$ and $\mu = L_{B,2}/L_{B,1}$ is the $B-$band luminosity ratio. Typically, $\mu \geq 1/4$ ($\Delta M_{B} \leq 1.5$) defines major mergers, and $\mu < 1/4$ defines minor mergers \cite[e.g.,][]{clsj11mmvvds}. We recall that $M_{B}^{\rm bri} = -22$ is assumed in the paper (Sect.~\ref{selection}). We note that Eq.~(\ref{Mselmask}) focus in a $B-$band luminosity selection, but the pair selection mask can be defined in the same way for mass-selected pairs if the stellar mass $M_{\star}(z)$ of the sources is known.

The $M_B\,(z)$ function of the principal galaxy and its companions in the ``system zero'' are shown in Fig.~\ref{ppf:3mb:fig}, and the derived luminosity ratios in Fig.~\ref{ppf:3Delta:fig}. We assumed $\mu = 1/4$ to select major companions. The first companion, {\it panel (a)} in both figures, is fainter than the central galaxy at each redshift with $\Delta M_B \sim 0.6$. This value is larger than the luminosity difference derived from the best-fitting solution, $\Delta M_B = 0.47$. We note that the principal galaxy fulfils the primary selection at $z > 0.52$. The second companion, {\it panel (b)} in both figures, is {\it brighter} than the principal galaxy at each redshift with $\Delta M_B \geq 1.2$, far from the best-fitting solution ratio, $\Delta M_B = 0.22$. The principal galaxy has a lower $\langle M_B \rangle$ than the companion, and was assumed as the central galaxy of the pair (Sect.~\ref{ppf:1}). However, the study of $M_B\,(z)$ reveals that the companion is indeed the central galaxy, and the principal galaxy of the ``system zero'' its satellite. As consequence, the companion has to fulfil the primary selection and the principal galaxy the secondary selection in Eq.~(\ref{Mselmask}). This case illustrates the possible complexity of the systems under study and the importance of analyse the physical variables in the redshift space.

\subsubsection{The pair probability function}\label{ppf}
At the end, each close pair system has associated a {\it pair probability function} defined as
\begin{equation}
{\rm PPF}(z) = \mathcal{Z}(z) \times \mathcal{M}^{\theta}(z) \times \mathcal{M}^{\rm pair}(z),
\end{equation}
where $\mathcal{Z}$ is the redshift probability function (Sect.~\ref{ppf:1}), $\mathcal{M}^{\theta}$ is the angular mask (Sect.~\ref{ppf:2}), and $\mathcal{M}^{\rm pair}$ is the pair selection mask (Sect.~\ref{ppf:2}) of the system. The PPF is a new probability function\footnote{Formally, the PPF {\it is not} a probability density function because its normalisation is different from one. However, the integral of the PPF {\it is a probability}. Thus, we keep the attribute probability in the following.} that encodes the relevant information about the close pairs in the survey. The PPFs are used to define the number of pairs (Sect.~\ref{ffpdf}), but are also capital for subsequent studies about the properties of galaxies with a close companion, such as the star formation rate. We will explore the potential of the PPFs in a future work.

\subsection{Correction by selection effects}\label{weights}
The PPFs defined in the previous section are mainly affected by two selection effects in ALHAMBRA: the selection in the \texttt{odds} parameter (Sec.~\ref{osr}) and the incompleteness in the search volume near the boundaries of the images (Sec.~\ref{border}). In the next sections, we define the statistical weights devoted to correcting for these selection effects.

\subsubsection{The \texttt{odds} sampling rate}\label{osr}
Following spectroscopic studies, we should correct the raw PPFs for the selection effects in our sample. As shown by \citet{molino13}, a selection in the $\mathcal{O}$ parameter ensures high quality photometric redshifts and a low rate of catastrophic outliers. \citet{clsj14ffcosvar} set $\mathcal{O} \geq 0.3$ as the optimal selection for merger fraction studies in ALHAMBRA. If the galaxies with $\mathcal{O} < 0.3$ are included in the samples, the projection effects become important and the merger fraction is overestimated. 

We define the {\it \texttt{odds} sampling rate} (OSR) as the ratio of galaxies with $\mathcal{O} \geq 0.3$ with respect to the total number of galaxies (i.e., those with $\mathcal{O} \geq 0$). The OSR mainly depends on the $I-$band magnitude because the quality of the photometric redshifts decrease according to the signal-to-noise. We use the redshift information encoded in the PDFs to estimate the OSR in our range of interest. Formally, the \texttt{odds} sampling rate of the ALHAMBRA sub-field $j$ is estimated as
\begin{equation}
{\rm OSR}_j = \frac{\sum_{i,\,\mathcal{O} \geq 0.3} \int_{z_{\rm min}}^{z_{\rm max}} {\rm PDF}_{i}\,(z)\,{\rm d}z}{\sum_{i,\, \mathcal{O} \geq 0} \int_{z_{\rm min}}^{z_{\rm max}} {\rm PDF}_{i}\,(z)\,{\rm d}z},
\end{equation}
where $i$ indexes every galaxy in the sub-field $j$.

The current ALHAMBRA release comprises 48 sub-fields. To compute the global OSR in ALHAMBRA, we combine the \texttt{odds} sampling rates of each sub-field $j$ as
\begin{equation}
{\rm OSR}_{\rm ALH} = \frac{\sum_{j} w^j_{\rm den} {\rm OSR}_j}{\sum_{j} w^j_{\rm den}},
\end{equation}
where $w^j_{\rm den}$ is the inverse of the number density in the field $j$. This density weight avoids the global OSR to be dominated by the ${\rm OSR}_j$ from the densest ALHAMBRA sub-fields.

\begin{figure}[t]
\centering
\resizebox{\hsize}{!}{\includegraphics{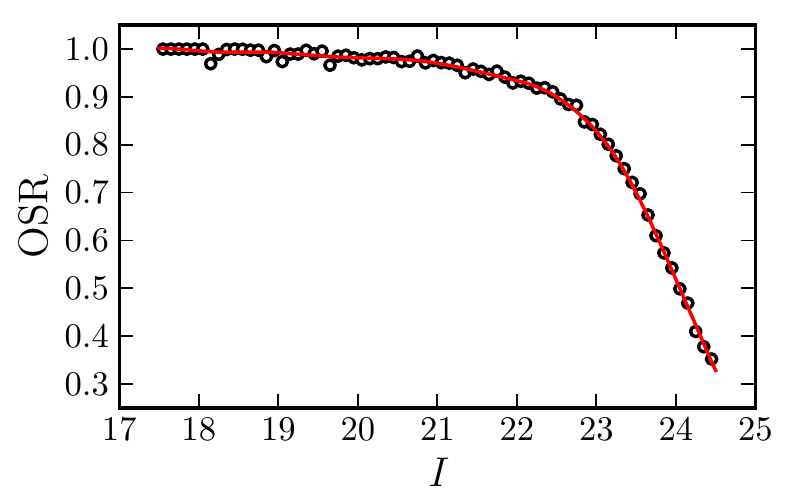}}
\caption{The odds sampling rate (OSR) in ALHAMBRA as a function of the $I-$band magnitude. Dots are the OSR estimated at $0.4 \leq z < 1$ in bins of 0.1 magnitudes. The solid line is the functional parametrisation of the OSR used in the paper. [{\it A colour version of this plot is available at the electronic edition}].}
\label{osrfig}
\end{figure}

We computed the ${\rm OSR}_{\rm ALH}$ in bins of 0.5 magnitudes in the $I$ band at $0.4 \leq z < 1$, and we interpolated the results to obtain ${\rm OSR}_{\rm ALH}\,(I)$. We checked that our interpolated function describes the ${\rm OSR}_{\rm ALH}$ properly estimating it in bins of 0.1 magnitudes (Fig.~\ref{osrfig}). Finally, we defined the \texttt{odds} weight of the galaxy $i$ as 
\begin{equation}
w_{\rm osr}^i = \frac{1}{{\rm OSR}_{\rm ALH}\,(I_i)}.
\end{equation}
The \texttt{odds} weight only depends on the $I-$band magnitude of the galaxy and we checked that it slightly depends on redshift in our range of interest.

\subsubsection{Border effects in the sky plane}\label{border}
When we search a primary source companion, we define a volume in the sky plane-redshift space. If the primary source is near the boundaries of the survey, a fraction of the search volume lies outside of the effective volume of the survey. We define the area weight of a close pair system as
\begin{equation}
w_{\rm area}(z) = \frac{1}{f_{\rm area}(z)},
\end{equation}
where $f_{\rm area}$ is the fraction of the search area that is covered by the ALHAMBRA survey. The search area is a ring centred at $(\alpha_1,\delta_1)$, and defined by $r_{\rm p}^{\rm min}$ and $r_{\rm p}^{\rm max}$. The search area, and therefore the area weight, depends on redshift because of the variation of the angular diameter distance $d_{\rm A}$ with $z$.

\subsubsection{The pair weight $w_{\rm pair}$}\label{wpair}
For each observed close pair we define the pair weight as
\begin{equation}
w_{\rm pair}(z) = w_{{\rm osr},1} \times w_{{\rm osr},2} \times w_{\rm area}(z),
\end{equation}
where $w_{{\rm osr},1} = w_{\rm osr}\,(I_1)$ is the \texttt{odds} weight of the central galaxy, $w_{{\rm osr},2} = w_{\rm osr}\,(I_2)$ is the \texttt{odds} weight of the satellite galaxy, and $w_{\rm area}$ is the area weight of the pair. The pair weight is always equal or larger than unity and it is applied to volume-limited samples.

\subsection{The merger fraction in photometric samples by PDF analysis}\label{ffpdf}
The merger fraction in the redshift range $z_{\rm r} = [z_{\rm min}, z_{\rm max})$ is
\begin{equation}
f_{{\rm m}} = \frac{\sum_k \int_{z_{\rm min}}^{z_{\rm max}} w_{\rm pair}^{k}(z) \times {\rm PPF}_k\,(z)\,{\rm d}z}{\sum_i \int_{z_{\rm min}}^{z_{\rm max}} w_{\rm osr}^i \times {\rm PDF}_i\,(z) \times \mathcal{S}_{1}^{i}\,(z) \,{\rm d}z} = \frac{\sum_k N_{\rm pair}^k}{\sum_i N_1^i},\label{ncphot}
\end{equation}
where $k$ indexes the close pair systems, $i$ indexes the galaxies in the primary sample, ${\rm PPF}$ is the pair probability function (Sect.~\ref{ppf}), $w_{\rm pair}$ is the pair weight (Sect.~\ref{wpair}), $w_{{\rm osr}}$ is the \texttt{odds} weight of the primary galaxies (Sect.~\ref{osr}), and $\mathcal{S}_1$ is the selection function of the primary galaxies (Sect.~\ref{selection}). Equation~(\ref{ncphot}) is the photometric analogous of Eq.~(\ref{ncspec}), with $\sum_k N_{\rm pair}^k$ being the number of close pairs and $\sum_i N_1^i$ the number of primary galaxies. In order to estimate the observational error of $f_{{\rm m}}$, as noted $\sigma_{f}$, we used the jackknife technique \citep{efron82}. We computed partial standard deviations for each system $k$, $\delta_k$, taking the difference between the measured $f_{{\rm m}}$ and the same quantity after removing the $k-$th pair from the sample, $f_{{\rm m}}^k$, such that $\delta_k = f_{{\rm m}} - f_{{\rm m}}^k$. For a redshift range with $N_{\rm p}$ systems, the variance is given by $\sigma_{f}^2 = [(N_{\rm p}-1) \sum_k \delta_k^2]/N_{\rm p}$.

\begin{table}
\caption{The ALHAMBRA close pair catalogue.}
\label{paircat}
\begin{center}
\begin{tabular}{ll}
\hline\hline\noalign{\smallskip}
\noalign{\smallskip}
Column & Description\\
\hline
\noalign{\smallskip}
PID		&	Identification number of the pair\\
ID1		&	ALHAMBRA ID of the principal galaxy\\
ID2		&	ALHAMBRA ID of the companion galaxy\\
RA1		&	Right ascension of the principal galaxy\\
DEC1		&	Declination of the principal galaxy\\
RA2		&	Right ascension of the companion galaxy\\
DEC2		&	Declination of the companion galaxy\\
theta		&	Angular separation (arcsec)\\
z1		&	Best photometric redshift of the principal galaxy\\
z2		&	Best photometric redshift of the companion galaxy\\
PPF		&	Integrated pair probability function\\
PPFw		&	Integrated PPF corrected by selection effects\\
I1		&	$F814W$ magnitude of the principal galaxy\\
I2		&	$F814W$ magnitude of the companion galaxy\\
wosr1		&	Odds weight of the principal galaxy\\
wosr2		&	Odds weight of the companion galaxy\\
warea		&	Average area weight of the pair\\
MB1		&	$M_B$ of the principal galaxy at z1\\
MB2		&	$M_B$ of the companion galaxy at z2\\
\noalign{\smallskip}
\hline
\end{tabular}
\end{center}
\end{table}

\subsection{Output of the code}\label{output}
\begin{figure}[t]
\centering
\resizebox{\hsize}{!}{\includegraphics{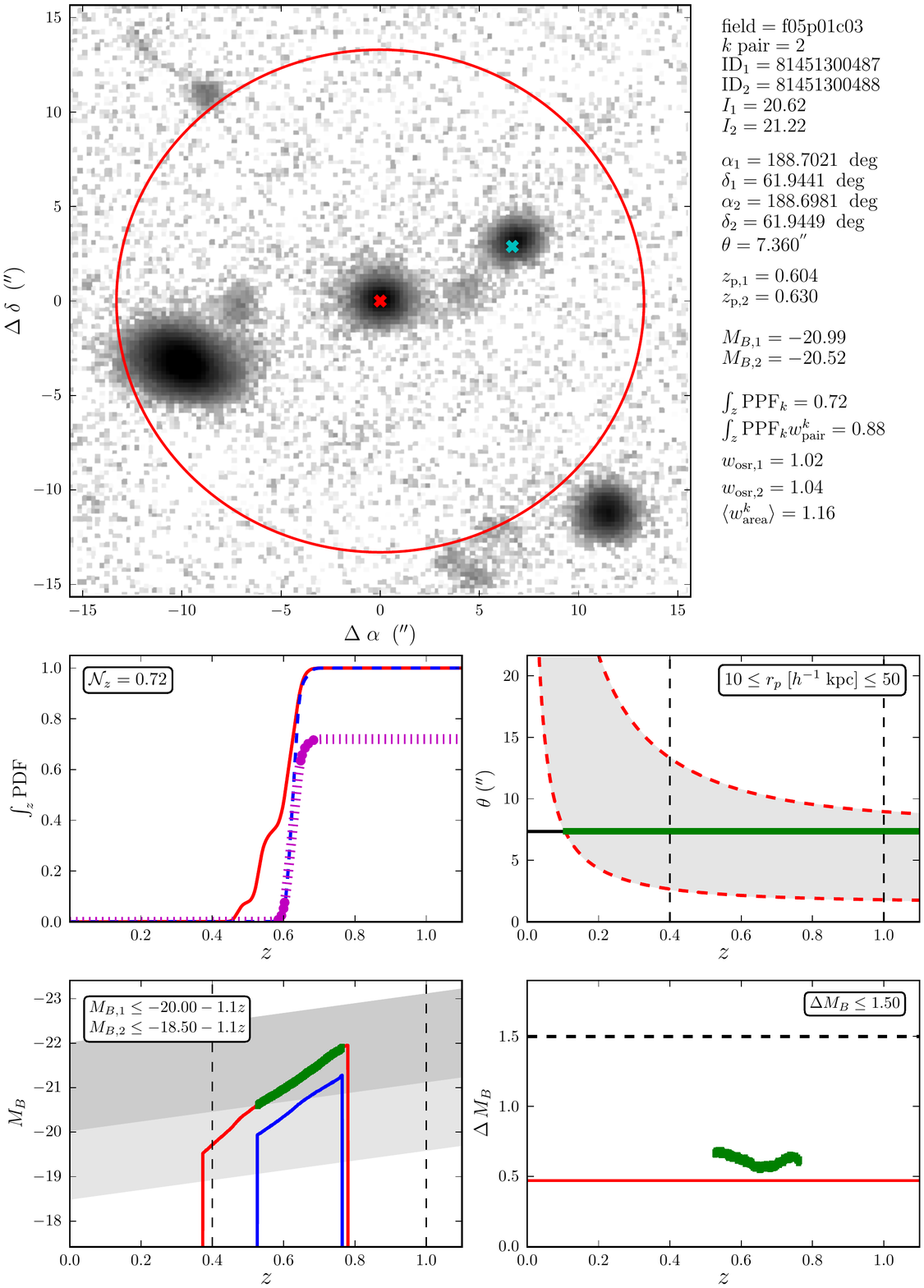}}
\caption{Example of the graphical output of the code. {\it Top panel}: Stamp of the close pair in the $I$ band. The red cross marks the principal galaxy and the blue one the companion galaxy. The red circle marks $r_p^{\rm max}$ at $z_{\rm min} = 0.4$. The text at the right summarises the main properties of the pair. {\it Middle--left panel}: The PDFs of the principal and the companion galaxy, and the function $\mathcal{Z}$ of the system (Sect.~\ref{ppf:1}, Fig.~\ref{ppf:1:fig}). {\it Middle--right panel}: The angular mask $\mathcal{M}^{\theta}$ of the system (Sect.~\ref{ppf:2}, Fig.~\ref{ppf:2:fig}). {\it Bottom panels} : The pair selection mask of the system. The {\it left panel} shows the selection of the primary and the secondary sample (Sect.~\ref{ppf:3}, Fig.~\ref{ppf:3mb:fig}), and the {\it right panel} shows the luminosity ratio constrain of the system (Sect.~\ref{ppf:3}, Fig.~\ref{ppf:3Delta:fig}). [{\it A colour version of this plot is available at the electronic edition}].}
\label{outpout}
\end{figure}

In addition to the merger fraction, the developed code also provides valuable outputs for future studies. The code creates three files:

\begin{itemize}
	\item {\it The close pair catalogue}. It summarises the main properties of the pairs, as shown in Table~\ref{paircat}. The reported values are either integrated over $z_{\rm min}$ and $z_{\rm max}$ or the values for the best photometric redshift. However, we encourage the use of the PDFs and the PPFs as outlined throughout the paper.
	\item {\it The close pair probabilities}. The relevant merger probabilities of the systems listed in the close pair catalogue are stored in a \texttt{hdf5} file. We report both the PPF and the $w_{\rm pair}$ of each close pair. The computation and the storage of the PPFs were done by \texttt{PyTables}\footnote{http://www.pytables.org/} \citep{pytables}.
	\item {\it A complete graphical output} with the summary of each close pair with $N_{\rm pair} \geq 0.01$. This summary includes the stamp of the merger system in the synthetic $I$ band, the relevant information from the close pair catalogue, the PDFs of the principal and companion galaxies, the function $\mathcal{Z}$, and both the angular and the pair selection masks. We present an example of the graphical output of the code in Fig.~\ref{outpout}.
\end{itemize}

The catalogues, probabilities, and figures of the ALHAMBRA major close pairs detected in Sect.~\ref{RMMsec} are available at https://cloud.iaa.csic.es/alhambra/catalogues/ClosePairs/

\section{The merger fraction in ALHAMBRA}\label{ffsec}

\subsection{A robust measurement of the merger fraction in ALHAMBRA}
As demonstrated by \citet{clsj14ffcosvar}, the 48 ALHAMBRA sub-fields can be assumed as independent for merger fraction studies. In addition, they set the optimal parameters to obtain reliable merger fractions. The ALHAMBRA merger fractions reported in the present paper are computed as follows:
\begin{enumerate}
\item The primary and secondary samples comprise those galaxies with $\mathcal{O} \geq 0.3$. This ensures high-quality photometric redshifts and non-biased samples, as shown by \citet{clsj14ffcosvar}.
\item The methodology presented in Sect.~\ref{metodo} was applied in each ALHAMBRA sub-field to obtain the merger fraction $f_{\rm m}$. This provided 48 estimations of $f_{\rm m}$ across the sky.
\item We applied the maximum likelihood estimator (MLE) presented in \citet{clsj14ffcosvar} to measure the average merger fraction in ALHAMBRA and its uncertainty. The MLE uses the measured merger fractions and their errors to compute the median of the merger fraction distribution. It also provides a reliable measurement of the intrinsic dispersion of the distribution, which is the cosmic variance. The cosmic variance for close pair studies in was studied in detail by \citet{clsj14ffcosvar}. We stress that the reported ALHAMBRA merger fractions are unaffected by cosmic variance.
\end{enumerate}

\begin{figure*}[th]
\centering
\resizebox{0.49\hsize}{!}{\includegraphics{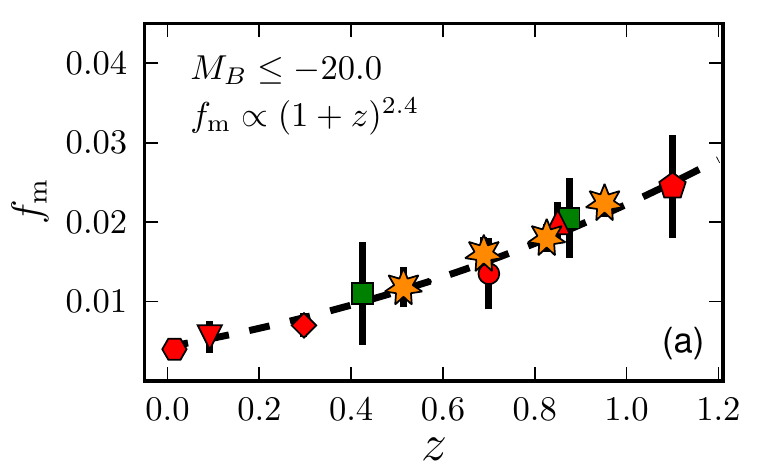}}
\resizebox{0.49\hsize}{!}{\includegraphics{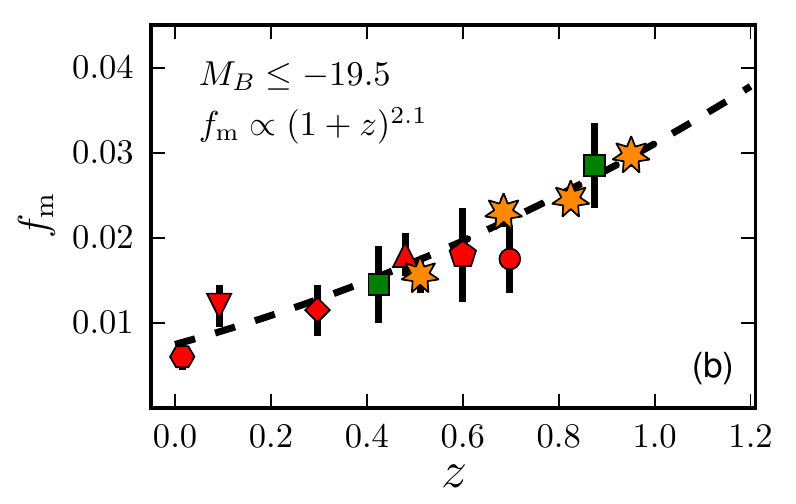}}
\resizebox{0.49\hsize}{!}{\includegraphics{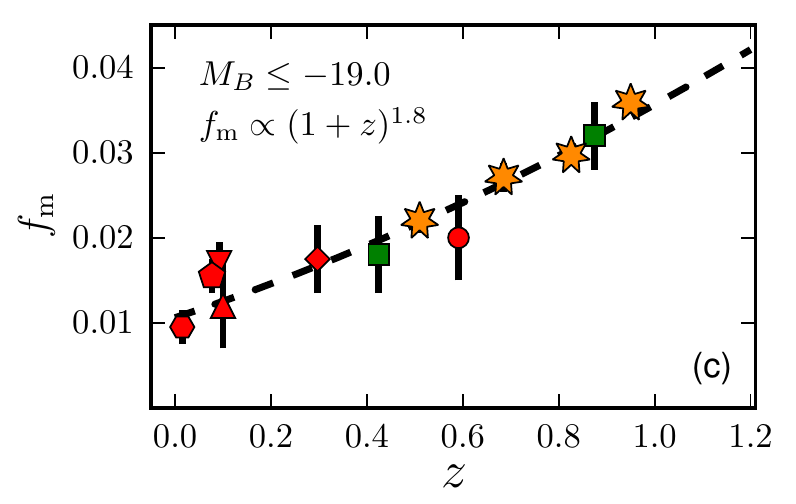}}
\resizebox{0.49\hsize}{!}{\includegraphics{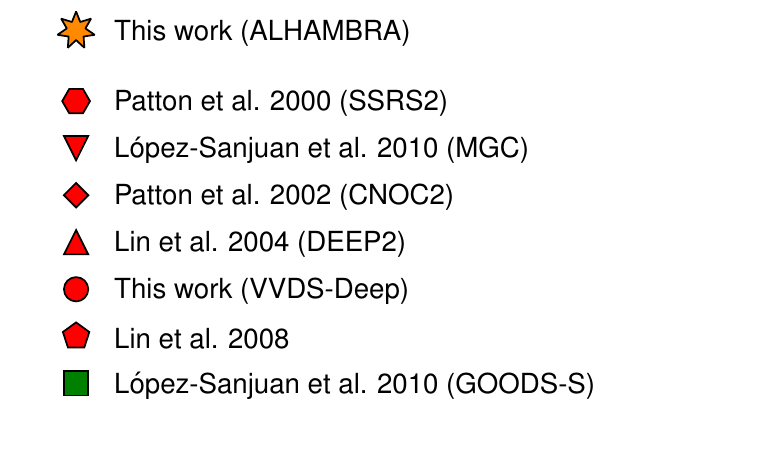}}
\caption{Merger fraction $f_{\rm m}$ as a function of redshift and the selection in $B$-band luminosity, {\it panel (a)} for $M_B \leq -20$ galaxies, {\it panel (b)} for $M_B \leq -19.5$ galaxies, and {\it panel (c)} for $M_B \leq -19$ galaxies. The orange stars are from the ALHAMBRA photometric survey (this work), the green squares from spectro-photometric pairs in GOODS-S (LS10), and the red symbols are from spectroscopic surveys: Hexagons from the SSRS2 \citep{patton00}, inverted triangles form the MGC (LS10), diamonds from the CNOC2 \citep{patton02}, dots from the VVDS-Deep (this work), triangles from the DEEP2 \citep{lin04}, and pentagons from \citet{lin08}. The dashed lines are the best fitting of Eq.~(\ref{ffzmb}) to the data. The power-law index from the best fitting is labelled in the panels. [{\it A colour version of this plot is available at the electronic edition}].}
\label{ff_vs_z_mb}
\end{figure*}

\begin{table*}
\caption{The merger fraction in ALHAMBRA as a function of the $B-$band luminosity.}
\label{ffmb}
\begin{center}
\begin{tabular}{lcccc}
\hline\hline\noalign{\smallskip}
Sample selection &        $z = 0.51$       &      $z = 0.69$        &        $z = 0.83$       &  $z = 0.95$\\
                 &   ($0.4 \leq z < 0.6$)  &  ($0.6 \leq z < 0.75$)  &  ($0.75 \leq z < 0.9$)  & ($0.9 \leq z < 1$)  \\
\noalign{\smallskip}
\hline
\noalign{\smallskip}
$M_B \leq -20$ 		& $0.0118 \pm 0.0025$ & $0.0160 \pm 0.0021$ & $0.0180 \pm 0.0018$ & $0.0224 \pm 0.0017$\\
$M_B \leq -19.5$ 	& $0.0156 \pm 0.0021$ & $0.0230 \pm 0.0017$ & $0.0245 \pm 0.0014$ & $0.0297 \pm 0.0014$\\
$M_B \leq -19$ 		& $0.0220 \pm 0.0015$ & $0.0271 \pm 0.0017$ & $0.0297 \pm 0.0013$ & $0.0359 \pm 0.0014$\\
\noalign{\smallskip}
\hline
\end{tabular}
\end{center}
\end{table*}

\subsection{The merger fraction in $M_{B}$ selected samples}
We test the reliability of our new methodology by comparing the merger fractions in the ALHAMBRA photometric survey with those from previous spectroscopic work. Robust measurements in the $B-$band from spectroscopic samples are available from the local Universe to $z \sim 1$, providing a valuable benchmark for our purposes. We used the homogenised compilation from LS10 to test the performance of the ALHAMBRA merger fractions. This compilation comprises the merger fractions from \citet{patton00} in the SSRS2 (Second Southern Sky Redshift, \citealt{ssrs2}) survey, LS10 in the MGC (Millennium Galaxy Catalogue, \citealt{mgc}; see also \citealt{depropris05,depropris07}) and GOODS-S (Great Observatories Origin Deep Survey South, \citealt{goods}), \citet{patton02} in the CNOC2 (Canadian Network for Observational Cosmology, \citealt{cnoc2}) survey, \citet{lin04} in the DEEP2 redshift survey \citep{deep2}, and \citet{lin08} in several of the above spectroscopic redshift surveys.

Following LS10, we defined three samples selected in $B-$band luminosity. These samples are defined with $M_{\rm B,1}^{\rm sel} = -20, -19.5$, and $-19$, and no evolution in the selection, $Q = 0$. We used these three samples as primary and secondary samples (i.e., $M_{\rm B,2}^{\rm sel} = M_{\rm B,1}^{\rm sel}$), and did not apply any luminosity condition between the galaxies in the pairs ($\mu = 0$). We searched close pairs with $6h^{-1}\ {\rm kpc} \leq r_{\rm p} \leq 21h^{-1}\ {\rm kpc}$ to mimic the definition used by LS10. We performed the study at $0.4 \leq z < 1$ to ensure large enough volumes at the lower redshifts and volume-limited samples at the higher ones. We summarise the ALHAMBRA merger fractions in Table~\ref{ffmb} and show them in Fig.~\ref{ff_vs_z_mb}. We find that the merger fraction increases with redshift and that it is larger for fainter samples.

LS10 report the number of companions $N_{\rm c}$, which is twice the number of close pairs (two galaxies per pair). We show $0.5 N_{\rm c}$ therefore in Fig.~\ref{ff_vs_z_mb}. In addition, we compute the merger fraction in the VVDS-Deep (VIMOS VLT Deep Survey, \citealt{lefevre05,vvdsud}) following Sect.~\ref{ffspec} and the completeness corrections outlined in \citet{deravel09} and \citet{clsj11mmvvds}. The ALHAMBRA merger fractions are in excellent agreement with the spectroscopic values. These results demonstrate that we can measure reliable and accurate merger fractions using photometric information only.

\subsection{The redshift evolution of the merger fraction in $M_{B}$ selected samples}
We parametrise the redshift evolution of the merger fraction with a power-law \cite[e.g.,][]{lefevre00},
\begin{equation}
f_{\rm m}\,(z) = f_{\rm m,0}\times(1+z)^{m}.
\end{equation}
Hereafter, the fittings are performed with \texttt{emcee} \citep{emcee}, a \texttt{Python} implementation of the affine-invariant ensemble sampler for Markov chain Monte Carlo (MCMC) proposed by \citet{goodman10}. \texttt{emcee} provides a collection of solutions in the parameters space, with the density of solutions being proportional to the posterior probability of the parameters. We obtain the best-fitting values and their uncertainties as the median and the dispersion of the projected solutions. In addition, the correlation between the parameters, as noted $\rho_{xy}$, is easily accessible.

\begin{figure}[t]
\centering
\resizebox{\hsize}{!}{\includegraphics{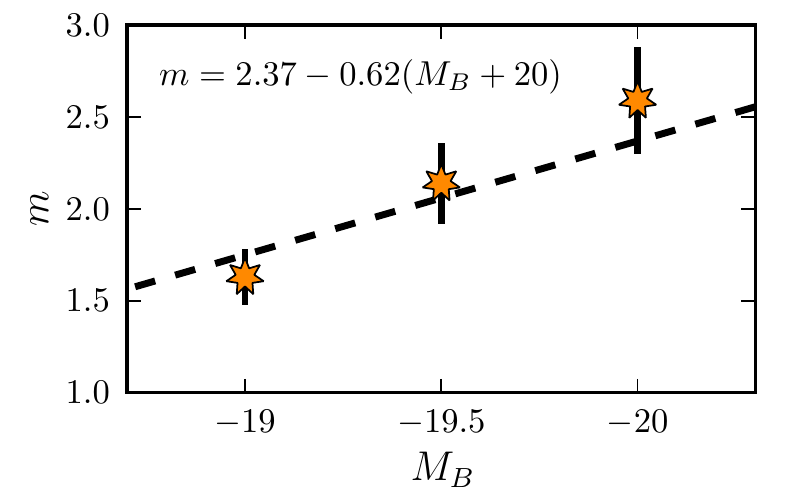}}\\
\resizebox{\hsize}{!}{\includegraphics{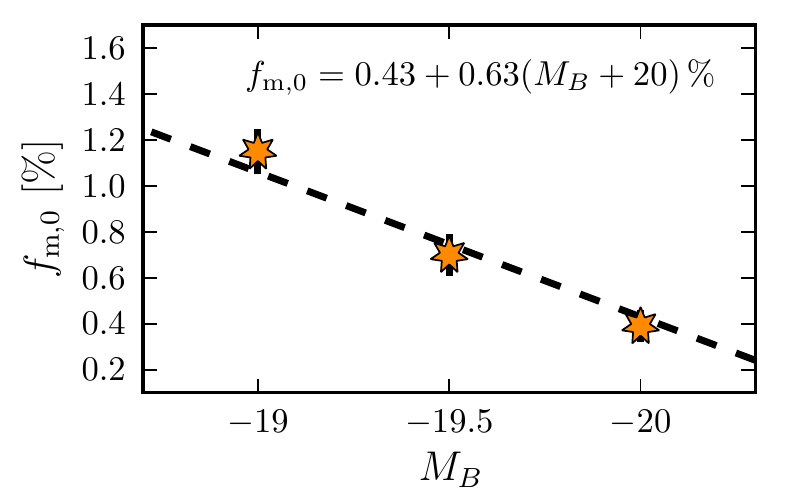}}
\caption{The power-law index ({\it top panel}) and the merger fraction at $z = 0$ ({\it bottom panel}) as a function of the $B-$band luminosity selection (Table~\ref{ffmbz}). The dashed lines are from the Eq.~(\ref{ffzmb}) fit to the data. [{\it A colour version of this plot is available at the electronic edition}].}
\label{m_vs_mb}
\end{figure}

We summarise the best fittings to the data from Fig.~\ref{ff_vs_z_mb} in Table~\ref{ffmbz}. We find that the power-law index $m$ increases with the luminosity selection, with the merger fraction at $z = 0$ decreasing (Fig.~\ref{m_vs_mb}). In addition, the parameters show a clear anti-correlation, with $\rho_{xy} \sim -0.96$ (Table~\ref{ffmbz}). 

We estimate the dependence of $f_{\rm m,0}$ and $m$ on the $B$-band luminosity selection by fitting the function
\begin{equation}
f_{\rm m}\,(z,M_B) = [f_{0} + \alpha(M_B + 20)]\times(1+z)^{m_0 + \beta(M_B + 20)}\label{ffzmb}
\end{equation}
to all the available data. We obtain $f_0 = 0.43 \pm 0.05\,$\%, $\alpha = 0.63 \pm 0.09\,$\%, $m_0 = 2.37 \pm 0.17$, and $\beta = -0.62 \pm 0.20$ (Fig.~\ref{m_vs_mb}). The individual parameters in Table~\ref{ffmbz} are compatible with this global fitting. These trends were already observed by \citet{lin04} and LS10, and they point out the importance of the selection when different merger fraction studies are compared.

\begin{table}
\caption{Redshift evolution of the merger fraction as a function of the $B-$band luminosity.}
\label{ffmbz}
\begin{center}
\begin{tabular}{lccc}
\hline\hline\noalign{\smallskip}
Sample selection &        $f_{{\rm m},0}$       &      $m$     &   $\rho_{xy}$\\
                 &        ($\%$)    &  &  \\
\noalign{\smallskip}
\hline
\noalign{\smallskip}
$M_B \leq -20$ 		& $0.39 \pm 0.07$ & $2.59 \pm 0.29$ & $-0.96$\\
$M_B \leq -19.5$ 	& $0.70 \pm 0.09$ & $2.14 \pm 0.22$ & $-0.97$\\
$M_B \leq -19$ 		& $1.15 \pm 0.10$ & $1.63 \pm 0.15$ & $-0.96$\\
\noalign{\smallskip}
\hline
\end{tabular}
\end{center}
\end{table}

\section{The major merger rate in ALHAMBRA}\label{RMMsec}
The final goal of merger studies is the estimation of the merger rate $R_{\rm m}$, defined as the number of mergers per galaxy and Gyr$^{-1}$. The merger rate is computed from the merger fraction by close pairs as
\begin{equation}
R_{\rm m} = C_{\rm p}\frac{C_{\rm m}}{T_{\rm m}}\,f_{\rm m},
\end{equation}
where the factor $C_{\rm p} = r_{\rm p}^{\rm max}/(r_{\rm p}^{\rm max} - r_{\rm p}^{\rm min})$ takes into account the lost companions at $r_{\rm p} < r_{\rm p}^{\rm min}$ \citep{bell06}, and $C_{\rm m}$ is the fraction of the observed close pairs that finally merge after a merger time scale $T_{\rm m}$. The merger time scale and the merger probability $C_{\rm m}$ should be estimated from simulations \citep[e.g.,][]{kit08,lotz10gas,lotz10t,lin10,jian12,moreno13}. On the one hand, $T_{\rm m}$ mainly depends on the search radius $r_{\rm p}^{\rm max}$, the stellar mass of the central galaxy, and the mass ratio between the galaxies in the pair with a mild dependence on redshift and environment \citep{kit08,jian12}. On the other hand, $C_{\rm m}$ mainly depends on $r_{\rm p}^{\rm max}$ and environment with a mild dependence on both redshift and the mass ratio between the galaxies in the pair \citep{jian12}. Despite the efforts in the literature to estimate both $T_{\rm m}$ and $C_{\rm m}$, different cosmological and galaxy formation models provide different values within a factor of two--three \citep[e.g.,][]{hopkins10mer}.

In the present paper the merger time scales from \citet{kit08} were used to translate our merger fractions and the merger fractions from the literature to a common scale. The $T_{\rm m}$ from \citet{kit08} already includes the merger probability, so we assume $C_{\rm m} = 1$ in the following.

\begin{table*}
\caption{The major merger rate of $M_{B} \leq -20 - 1.1z$ galaxies in ALHAMBRA.}
\label{RMM}
\begin{center}
\begin{tabular}{lcccc}
\hline\hline\noalign{\smallskip}
Sample selection &        $z = 0.51$       &      $z = 0.68$        &        $z = 0.82$       &  $z = 0.95$\\
                 &   ($0.4 \leq z < 0.6$)  &  ($0.6 \leq z < 0.75$)  &  ($0.75 \leq z < 0.9$)  & ($0.9 \leq z < 1$)  \\
\noalign{\smallskip}
\hline
\noalign{\smallskip}
Full sample		& $0.058 \pm 0.009$ & $0.071 \pm 0.011$ & $0.086 \pm 0.012$ & $0.092 \pm 0.012$\\
Red galaxies (E/S0) 	& $0.078 \pm 0.014$ & $0.097 \pm 0.018$ & $0.108 \pm 0.018$ & $0.104 \pm 0.019$\\
Blue galaxies (S/SB) 	& $0.041 \pm 0.006$ & $0.048 \pm 0.007$ & $0.067 \pm 0.009$ & $0.074 \pm 0.009$\\
\noalign{\smallskip}
\hline
\end{tabular}
\end{center}
\end{table*}

\subsection{The major merger rate of bright galaxies}
In this section, we estimate the major merger rate $R_{\rm MM}$ of bright galaxies in the ALHAMBRA survey and we compare it with data from the literature. We define primary galaxies with $M_{B,1}^{\rm sel} = -20$, taking $Q = 1.1$ as the evolution of the luminosity function with $z$ \citep[e.g.,][]{ilbert06}. This selects galaxies brighter than $L^*_{B}$ up to $z = 1$. We searched major companions with $\Delta M_B \leq 1.5$ magnitudes ($\mu \geq 1/4$). The companion sample therefore comprises galaxies with $M_{B,2}^{\rm sel} = -18.5$ and $Q = 1.1$. 

\begin{figure}[t]
\centering
\resizebox{\hsize}{!}{\includegraphics{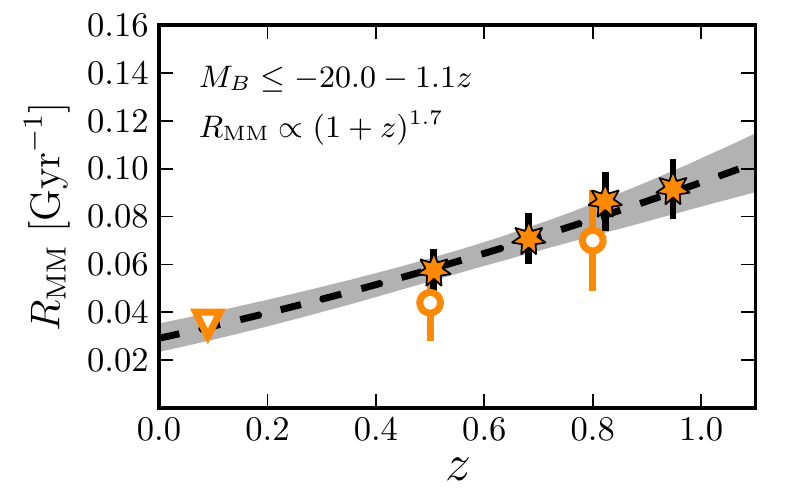}}
\caption{Major merger rate $R_{\rm MM}$ as a function of redshift for $M_{B} \leq -20 - 1.1z$ galaxies. The stars are from the ALHAMBRA photometric survey, the circles from the VVDS-Deep spectroscopic survey, and the inverted triangle from the MGC spectroscopic survey. The dashed line is the best fitting of a power-law to the data. The power-law index of the best fitting is labelled in the panel. The gray area marks the 68\% confidence interval of the fitting. [{\it A colour version of this plot is available at the electronic edition}].}
\label{RMM_vs_z}
\end{figure}

We estimated the major merger rate from the merger fraction of $10h^{-1}\ {\rm kpc} \leq r_{\rm p} \leq 50h^{-1}$ kpc close pairs, and following \citet{clsj11mmvvds} we used $T_{\rm m} = 2.3 \pm 0.3$ Gyr. We summarise the ALHAMBRA merger rates in Table~\ref{RMM} and in Fig.~\ref{RMM_vs_z}. We find that the major merger rate increases with redshift, in agreement with previous work \citep[e.g.,][]{deravel09}. We test the robustness of the ALHAMBRA results by comparing them with those from the VVDS-Deep and MGC spectroscopic surveys \citep{clsj11mmvvds}. We find that the ALHAMBRA data agree with the major merger rates from the VVDS-Deep at $0.3 < z < 1$. We parametrise the major merger rate as
\begin{equation}
R_{\rm MM}\,(z) = R_{\rm MM,0}\times(1+z)^{n}.
\end{equation}
The best fitting to the ALHAMBRA, VVDS-Deep, and MGC data is presented in Table~\ref{RMMz}. We find $R_{\rm MM,0} = 0.029 \pm 0.006$ Gyr$^{-1}$ and $n = 1.69 \pm 0.37$. These values from the combined data set are consistent with those obtained from the ALHAMBRA data alone, with $R_{\rm MM,0} = 0.032 \pm 0.013$ Gyr$^{-1}$ and $n = 1.57 \pm 0.74$. These results further support our new methodology and the quality of the ALHAMBRA survey data.

\begin{figure}[t]
\centering
\resizebox{\hsize}{!}{\includegraphics{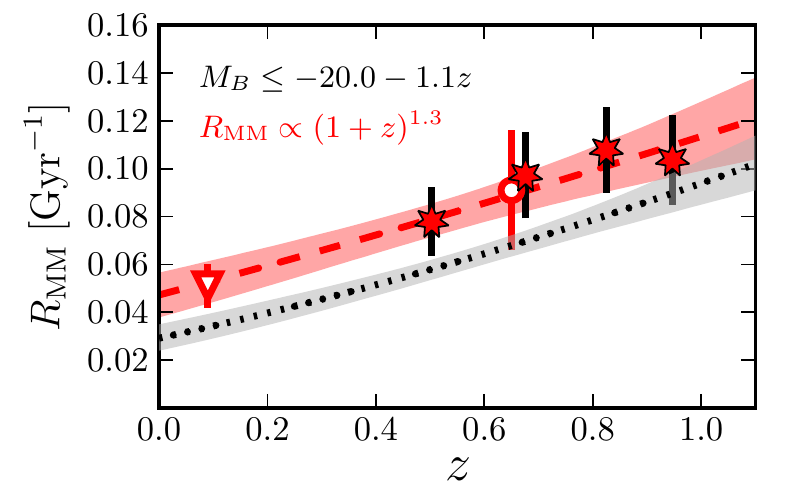}}
\resizebox{\hsize}{!}{\includegraphics{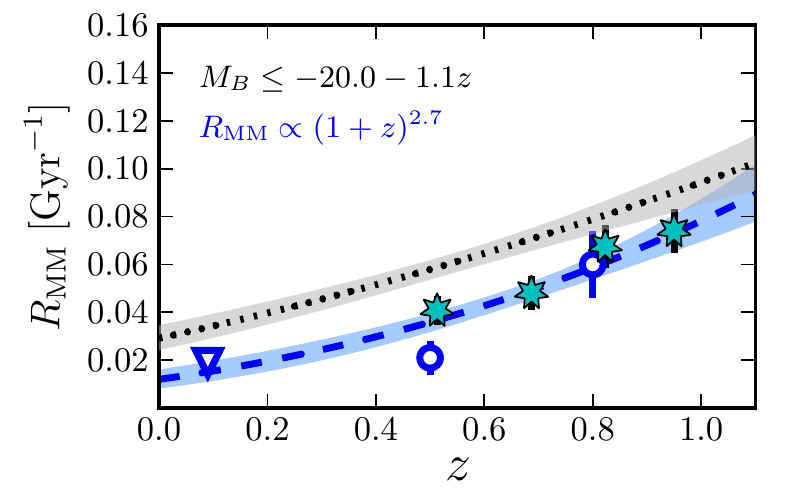}}
\caption{Major merger rate $R_{\rm MM}$ as a function of redshift and colour for $M_{B} \leq -20 - 1.1z$ galaxies. The stars are from the ALHAMBRA photometric survey, the circles from the VVDS-Deep spectroscopic survey, and the inverted triangles from the MGC spectroscopic survey. The dashed line in both panels is the best fitting of a power-law to the data. The power-law index of the best fitting is labelled in the panels. The coloured area marks the 68\% confidence interval of the fitting. The dotted lines and the grey areas mark the best fitting to the global population shown in Fig.~\ref{RMM_vs_z}. {\it Top panel}: Red population, selected as galaxies with E/S0 templates in ALHAMBRA, galaxies with $NUV-r \geq 4.25$ in the VVDS-Deep, and galaxies with $u-r \geq 2.1$ in the MGC. {\it Bottom panel}: Blue population, selected as galaxies with S/SB templates in ALHAMBRA, galaxies with $NUV-r < 4.25$ in the VVDS-Deep, and galaxies with $u-r < 2.1$ in the MGC. [{\it A colour version of this plot is available at the electronic edition}].}
\label{RMMcol_vs_z}
\end{figure}

\begin{figure}[t]
\centering
\resizebox{\hsize}{!}{\includegraphics{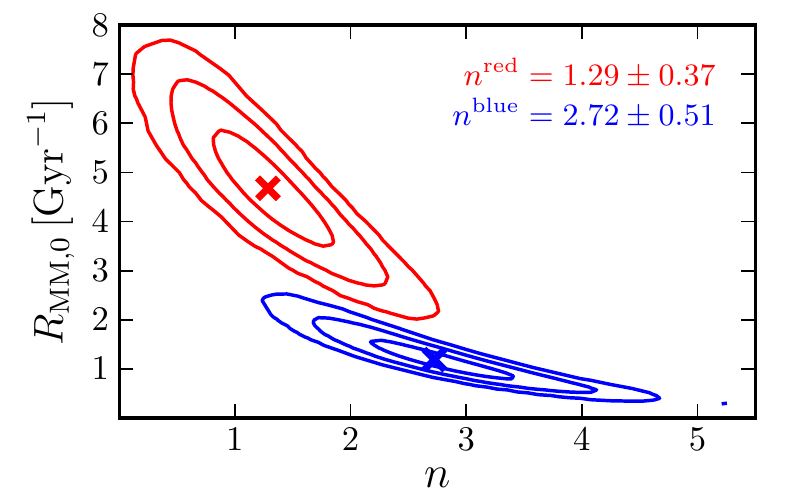}}
\caption{Probability contours in the $R_{\rm MM,0}$ vs $n$ plane for red and blue galaxies. The contours enclose 68.2\%, 95.4\% and 99.7\% of the probability. The crosses mark the most probable values of the parameters. The best power-law indices $n$ are labelled in the panel. [{\it A colour version of this plot is available at the electronic edition}].}
\label{RMM_vs_n}
\end{figure}

\subsection{The major merger rate of red and blue galaxies}
In this section we study the major merger rate of red (E/S0 templates) and blue (S/SB templates) galaxies. The primary and the secondary samples are defined as in the previous section. We estimated the major merger rate of red galaxies using the ${\rm PDF}^{\rm red}$ and of blue galaxies using the ${\rm PDF}^{\rm blue}$. As noted in Sect.~\ref{pdfs}, we did not perform any colour selection of the sources and all the galaxies in the primary sample are included in the analysis. This is a novel approach only possible thanks to the rich information encoded in the PDFs provided by BPZ2.0. We used the full PDFs of the companions, i.e., we looked for all the possible companions of red and blue primary galaxies. Following \citet{clsj11mmvvds}, we used $T_{\rm m}^{\rm red} = 2.1 \pm 0.3$ Gyr for red galaxies and $T_{\rm m}^{\rm blue} = 2.6 \pm 0.3$ Gyr for blue galaxies (i.e., the blue galaxies are less massive than the red galaxies of similar $B-$band luminosity).

We summarise our results in Table~\ref{RMM} and Fig.~\ref{RMMcol_vs_z}. We find that the major merger rate of red galaxies is larger than the major merger rate of blue galaxies at any redshift. As in the previous section, we compare the ALHAMBRA results with those from the VVDS-Deep and the MGC. \citet{clsj11mmvvds} define red galaxies in the VVDS-Deep with $NUV-r \geq 4.25$ and blue galaxies with $NUV-r < 4.25$. In addition, we computed the red and blue merger rates in the MGC. We defined red galaxies with $u-r \geq 2.1$ and blue galaxies with $u-r < 2.1$ \citep[e.g.,][]{strateva01} thanks to the SDSS (Sloan Digital Sky Survey, \citealt{sdssdr8}) photometry. We find $R_{\rm MM, MGC}^{\rm red} = 0.051 \pm 0.009$ Gyr$^{-1}$ and $R_{\rm MM, MGC}^{\rm blue} = 0.019 \pm 0.005$ Gyr$^{-1}$ at $z = 0.09$. The ALHAMBRA major merger rates are in agreement with the spectroscopic values.

We fit a power-law to the data and we find that (Table~\ref{RMMz})
\begin{itemize}
\item The evolution of the red merger rate is
	\begin{equation}
	R_{\rm MM}^{\rm red} = (0.047 \pm 0.008) \times (1+z)^{1.29 \pm 0.37}\,{\rm Gyr}^{-1}.
	\end{equation}

\item The evolution of the blue merger rate is
	\begin{equation}
	R_{\rm MM}^{\rm blue} = (0.012 \pm 0.003) \times (1+z)^{2.72 \pm 0.51}\,{\rm Gyr}^{-1}.
	\end{equation}
\item The blue merger rate evolves faster than the red merger rate, in agreement with previous results \citep[e.g.,][]{lin08,deravel09,chou10,clsj12sizecos}.
\end{itemize}

We note that the parameters in the fittings are anti-correlated (Table~\ref{RMM}). To illustrate this correlation, we show the probability contours of the fitted parameters in the Fig.~\ref{RMM_vs_n}. This figure demonstrates that the fittings to the red and the blue populations are different at more than $3\sigma$, even if the indices $n$ are compatible at $2\sigma$ level. This anti-correlation has also an impact in the integrated merger history of red and blue galaxies, which is much better constrained than the individual parameters from the fitting. Integrating the merger rate over cosmic time, we find that the average number of mergers per galaxy since $z = 1$ is $N_{\rm m}^{\rm red} = 0.57 \pm 0.05$ for red galaxies and $N_{\rm m}^{\rm blue} = 0.26 \pm 0.02$ for blue galaxies. Thus, red galaxies have undergone $\sim2$ times more major mergers than blue galaxies since $z = 1$.

These results demonstrate that our new methodology deals naturally with colour segregations and that accurate merger rates of red and blue galaxies can be estimated with only photometric data. 

\begin{table}
\caption{The major merger rate evolution of $M_{B} \leq -20 - 1.1z$ galaxies.}
\label{RMMz}
\begin{center}
\begin{tabular}{lccc}
\hline\hline\noalign{\smallskip}
Sample selection &        $R_{{\rm MM},0}$       &      $n$   &      $\rho_{xy}$\\
                 &        (Gyr$^{-1}$)    &    &  \\
\noalign{\smallskip}
\hline
\noalign{\smallskip}
Full sample		& $0.029 \pm 0.006$ & $1.69 \pm 0.36$ & $-0.93$ \\
Red galaxies (E/S0) 	& $0.047 \pm 0.008$ & $1.29 \pm 0.37$ & $-0.91$ \\
Blue galaxies (S/SB) 	& $0.012 \pm 0.003$ & $2.72 \pm 0.51$ & $-0.96$ \\
\noalign{\smallskip}
\hline
\end{tabular}
\end{center}
\end{table}

\section{Summary and conclusions}\label{conclusions}

We have developed a new methodology to compute accurate merger fractions by PDF analysis of photometric close pairs. Our method solves the main shortcomings present in previous merger fraction studies in photometric samples by (i) using the full PDF of the sources in redshift space, (ii) including the variation in the luminosity of individual sources with $z$ in both the selection of the samples and in the luminosity ratio constrain, and (iii) splitting individual PDFs in red and blue spectral templates to deal robustly with rest-frame colour selections.

We find that our methodology provides merger fractions and rates in nice agreement with those from spectroscopic work, both for the general population and for red and blue galaxies. With the merger rate of bright ($M_B \leq -20 - 1.1z$) galaxies evolving as $(1+z)^n$, the power-law index $n$ is larger for blue galaxies ($n = 2.7\pm0.5$) than for red galaxies ($n = 1.3\pm0.4$), confirming previous results. Integrating the merger rate over cosmic time, we find that the average number of mergers per galaxy since $z = 1$ is $N_{\rm m}^{\rm red} = 0.57 \pm 0.05$ for red galaxies and $N_{\rm m}^{\rm blue} = 0.26 \pm 0.02$ for blue galaxies. Thus, red galaxies have undergone $\sim2$ times more major mergers than blue galaxies since $z = 1$.

We conclude that our new methodology provides accurate merger fractions from photometric data alone, dealing with the available information in both redshift and template spaces robustly. We have tested the performance of our new methodology with the PDFs provided by the ALHAMBRA survey, but it can be applied to any current and future photometric survey, such as DES, J-PAS, or LSST.

In future work we will study the dependence of the merger fraction on the stellar mass or the morphology (see \citealt{povic13}, for
details about the morphological classification in ALHAMBRA). In addition, the study of galaxy properties in paired galaxies will be performed thanks to the PPFs defined in the present work. Finally, the comparison of the observed trends with the expectations from cosmological models should be explored to better understand the role of mergers in galaxy evolution.

\begin{acknowledgements}
We dedicate this paper to the memory of our six IAC colleagues and friends who
met with a fatal accident in Piedra de los Cochinos, Tenerife, in February 2007,
with a special thanks to Maurizio Panniello, whose teachings of \texttt{python}
were so important for this paper.\\
This work has been mainly funding by the FITE (Fondos de Inversiones de Teruel) and the projects AYA2012-30789, AYA2006-14056, and CSD2007-00060. We also acknowledge the financial support from the Spanish Government grants AYA2010-15169, AYA2010-22111-C03-01, AYA2010-22111-C03-02, and AYA2013-48623-C2-2, from the Junta de Andaluc\'{\i}a through TIC-114 and the Excellence Project P08-TIC-03531, and from the Generalitat Valenciana through the projects Prometeo/2009/064 and PrometeoII/2014/060.\\
A.~J.~C. is {\it Ram\'on y Cajal} fellow of the Spanish government.\\
This research made use of \texttt{Astropy}, a community-developed core \texttt{Python} package for Astronomy \citep{astropy}, and \texttt{Matplotlib}, a 2D graphics package used for \texttt{Python} for publication-quality image generation across user interfaces and operating systems \citep{pylab}.

\end{acknowledgements}

\bibliography{biblio}
\bibliographystyle{aa}

\end{document}